\documentclass[12pt]{article}

\newlength{\dinwidth}                                                                         
\newlength{\dinmargin}                                                                         
\def\lapproxeq{\lower .7ex\hbox{$\;\stackrel{\textstyle                                                                         
<}{\sim}\;$}}                                                                         
\def\gapproxeq{\lower .7ex\hbox{$\;\stackrel{\textstyle                                                                         
>}{\sim}\;$}}        
\usepackage{amssymb}
\usepackage{amsmath}
\usepackage{epsfig}
\usepackage{graphicx}
\usepackage{color}
\usepackage{physics}
\usepackage{siunitx}
\usepackage{epstopdf}
\usepackage{hyperref}

\newcommand{\dipoleS}[1][\xtarget]{S^{(2)}_{#1}}
\newcommand{\dipoleSbare}{S^{\star(2)}}
\newcommand{\dipoleSadj}[1][\xtarget]{\tilde{S}^{(2)}_{#1}}
\newcommand{\quadrupoleS}[1][\xtarget]{S^{(4)}_{#1}}
\def\pperp{p_{\perp}}
\def\kperp{k_{\perp}}
\def\kgperp{k_{g\perp}}
\def\xperp{x_{\perp}}
\def\yperp{y_{\perp}}
\def\bperp{b_{\perp}}

\def\rperp{r_{\perp}}
\def\alphas{\alpha_s}
\def\coherencetime{t_c}
\def\targettime{\tau}
\def\hadronrapidity{y_h}
\def\fragfrac{z}
\def\emitfrac{\xi}
\def\facscale{\mu}

\def\qperp{q_{\perp}}

\def\uperp{u_\perp}

\def\Nc{N_c}
\def\CF{C_F}

\def\xprojectile{x_p}
\def\xtarget{x_g}
\def\gluonrapidity{y_g}
\newcommand{\dipoleF}[1][\xtarget]{\mathcal{F}_{#1}}
\newcommand{\dipoleFbare}{\dipoleF[]^{\star}}
\newcommand{\dipoleFadj}[1][\xtarget]{\tilde{\mathcal{F}}_{#1}}

\def\besselJzero{J_0}

\def\satscale{Q_s}
\def\sqs{\sqrt{s}}
\def\sqsnn{\sqrt{s_{NN}}}
\def\rapidity{Y}
\def\quarkrapidity{y}

\def\rapfac{\xi_{\mathrm{f}}}
\newcommand{\partondistLO}[2]{\ensuremath{#2 #1\left(#2\right)}}
\newcommand{\partondist}[3]{\ensuremath{#2 #1\left(#2, #3\right)}}
\newcommand{\fragmentationpurefunc}[2]{\ensuremath{D_{#2/#1}}}
\newcommand{\fragmentationLO}[3]{\ensuremath{\fragmentationpurefunc{#1}{#2}\left(#3\right)}}
\newcommand{\fragmentation}[4]{\ensuremath{\fragmentationpurefunc{#1}{#2}\left(#3, #4\right)}}

\def\bjorkenx{x}
\def\mandelstams{s}
\def\mandelstamt{t}

\def\massnumber{A}
\def\probability{P}
\def\bfklkernel{K}
\def\bkkernel{\mathcal{K}}
\def\dipoleN{N}
\def\evolutionrapidity{Y}
\def\alphasbar{\expandafter\bar\alphas}
\def\gluonsplitting{x}
\def\Uwilson{U}
\def\Wwilson{W}
\def\pathordered{\mathcal{P}}
\def\strongcoupling{g_s}
\def\defn{\equiv} 
\def\quarkxsec{\sigma^{pA \to qX}}
\def\sigmainclusive{\sigma^{pA \to hX}}
\def\sigmainclusivepp{\sigma^{pp \to hX}}
\def\LambdaQCD{\Lambda_\text{QCD}}
\def\sqs{\sqrt{s}}
\def\xglo{x_{g0}}
\def\rapidity{Y}
\def\Ncoll{N_{\text{coll}}}
\def\pA{\ensuremath{\mathrm{p}A}}
\newcommand\msbar{\ensuremath{\overline{\mathrm{MS}}}}
\newcommand\h[4]{\mathcal{H}_{#2#3#4}^{(#1)}}
\newcommand{\beeq}{\begin{eqnarray}}
\newcommand{\eeeq}{\end{eqnarray}}
\newcommand{\be}{\begin{equation}}
\newcommand{\ee}{\end{equation}}
\newcommand{\bea}{\begin{array}}
\newcommand{\eea}{\end{array}}
\let\oldvec\vec
\def\vec#1{\expandafter\oldvec#1}
\definecolor{dzcolor}{rgb}{0,0.1,0.7}
\definecolor{ascolor}{rgb}{1,0,1}

\newcommand\pkout{\marginpar{\color{dzcolor}$\clubsuit$}\bgroup\markoverwith{\color{dzcolor}{\rule[0.4ex]{2pt}{0.8pt}}}\ULon}
\newcommand\asout{\marginpar{\color{ascolor}$\heartsuit$}\bgroup\markoverwith{\color{ascolor}{\rule[0.4ex]{2pt}{0.8pt}}}\ULon}

\begin{document}

%
%
\titlepage
\vspace*{2cm}
\begin{center}
{\Large \bf Saturation in inclusive production beyond leading logarithm accuracy}

\vspace*{1cm}
{Anna M. Sta\'sto$^{a}$ and David Zaslavsky$^{b}$}

\vspace*{0.5cm}
$^a$ {\it Physics Department, Pennsylvania State University, 104 Davey Laboratory\\University Park, Pennsylvania 16802, USA\\}{\small astasto@phys.psu.edu}

\vspace*{0.2cm}
$^b$ {\it Institute of Particle Physics, Central China Normal University\\Wuhan, Hubei, China\\}{\small david.zaslavsky@mailaps.org}
\end{center}

\vspace*{2cm}

\begin{abstract}
We review  recent progress on the calculations on the inclusive forward hadron production within the saturation formalism. After introducing the concept of perturbative parton saturation and nonlinear evolution we discuss the formalism for the forward hadron production at high energy in the leading and next-to-leading order. Numerical results are presented and compared with the experimental data on forward hadron production in $dA$ and $pA$. We discuss the problem of the negativity of the NLO cross section at high transverse momenta, study its origin in detail and present possible improvements which include the corrected kinematics and the suitable choice of the rapidity cutoff. 
\end{abstract}
\newpage
\section{Introduction}
\label{sec:introduction}

Modern particle colliders like the Relativistic Heavy Ion Collider (RHIC) and the Large Hadron Collider (LHC) allow physicist to probe the dynamics of strongly interacting matter under extreme conditions of high energy and density.  One of the main goals of these experiments is to create and test the new form of strongly interacting matter, the quark-gluon plasma. A plethora of experimental data, for example on jet quenching and the elliptic flow~\cite{Arsene:2004fa,Back:2004je,Adams:2005dq,Adcox:2004mh,ATLAS:2011ah,ATLAS:2012at,Aamodt:2010pa} confirm the existence of this strongly coupled system with collective effects, whose subsequent evolution is tractable within hydrodynamical approaches~\cite{Kolb:2003dz}.

One of the main questions that arise, when trying to describe the properties and dynamics of such a complex system, is the role of the initial state in the highly energetic collisions of hadrons and nuclei. The standard approaches to describe the wide variety of processes in hadronic collisions are based on the collinear approximation~\cite{Collins:1989gx} and are grounded on the assumption of the presence of high momentum scales, which justifies the use of perturbation theory and allows for the factorization of the cross section into perturbative matrix elements, which are process-specific, and universal parton distributions and fragmentation functions. This framework is highly successful in the description of phenomena that involve large scales and relatively small numbers of partons.

On the other hand, it has been well known since the deep inelastic scattering experiments at the HERA collider that at high energy, or equivalently at small values of Bjorken $\bjorkenx$, the structure function of the proton grows very fast with decreasing values of $\bjorkenx$. This is related to the self-interaction of gluons in QCD, which leads to a strong increase of the density of gluons in the low-$\bjorkenx$ regime, and subsequently of the observable cross sections in deep inelastic scattering. The famous Balitsky-Fadin-Kuraev-Lipatov (BFKL) \cite{Balitsky:1978ic,Kuraev:1977fs} evolution equation, which sums gluon emissions in the Regge limit --- that is, when $\mandelstams \gg \abs{\mandelstamt}$ and $\alphas \ln\mandelstams$ is large --- leads to strong power growth of the scattering amplitude. This multiplication of gluons with increasing energy can eventually lead to a regime where the densities are high and a new computational framework has to be developed in order to account for the complexities of the multiparton state.

Various calculations within Quantum Chromodynamics predict that in such a case, the phenomenon of parton saturation occurs~\cite{Gribov:1984tu,Mueller:1985wy}: in addition to the gluon splitting which leads to the growth of the density, a competing process emerges which tames the growth of the gluon densities and consequently of the observable cross section. The onset of this behavior depends not only on the energy of the process in question, but also on the type of the particle involved in the collision. By changing the size of the initial particle from a proton to a heavy nucleus, the process of gluon recombination is enhanced by a factor proportional to a certain power of the mass number, and the energy at which it can happen is lowered with respect to the smaller particle. The calculations in the high-energy and high-density limit predict the emergence of an energy-dependent scale which characterizes the onset of parton saturation, namely the saturation scale $\satscale(\xtarget, \massnumber)$. It is predicted that this scale increases with the energy and the mass number $\massnumber$, though its absolute normalization is not yet predicted by these calculations, and has to be deduced from comparisons with the experimental data.
In that context, knowledge of the partonic initial state and the possible onset of the parton saturation is essential for the complete description of both proton-proton as well as proton-ion collisions, and it will shed more light into the initial state for the heavy ion collisions.

There have been numerous tests of parton saturation in a variety of processes which involve both protons and nuclei. Among them are proton-nucleus collisions, which serve as one of the important benchmark processes for the heavy ion collisions. For a review see e.g.~\cite{Albacete:2013tpa}. In a proton-nucleus collision, the proton is relatively dilute and is probing the dense system of the heavy nucleus.

Of particular importance is the study of the forward single inclusive production of hadrons in proton (or deuteron) collisions with nuclei. By observing the hadron which is produced in the forward direction of the proton (or deuteron), the kinematics are such that the participating parton from the proton side has a large value of fractional momentum $\xprojectile$, and as such is usually a valence parton. On the other hand, the parton from the nucleus side has a small value of $\xtarget \ll 1$. Therefore, the process is sensitive to the nuclear parton density at small $\bjorkenx$, and is dominated by the gluon density.

It was observed, in the data from BRAHMS at the RHIC, that the ratio of the single inclusive production in deuteron-nucleus collisions to that in proton-proton collisions was suppressed at forward rapidity. This phenomenon had a natural explanation within the parton saturation framework, which predicted enhanced suppression with increasing rapidity~\cite{Kharzeev:2003wz,Kharzeev:2004yx}. Nevertheless, the interpretation of these data was not unique, as the process occurred near the kinematic boundary, and other effects not necessary related to  the parton saturation could also be responsible for the observed suppression.

On the other hand, the accuracy of calculations which include parton saturation in the description of the experimental data, from deep inelastic scattering to nuclear collisions, still has to be better quantified by incorporating higher order corrections. Most of the calculations so far, which were compared to experimental data, were performed at leading logarithmic order in $\ln 1/\bjorkenx$, sometimes including only part of the next-to-leading order corrections; for example, in the form of the running coupling in the nonlinear evolution equations. Clearly, in order to make more robust predictions with smaller theoretical uncertainties, one needs to go beyond the lowest order calculations.  During the last decade, there have been a number of theoretical derivations which aim to go beyond the lowest order approximation in the small-$\bjorkenx$ limit, for example the derivation of the next-to-leading order corrections to the nonlinear evolution equations~\cite{Camici:1997ij,Ciafaloni:1998gs,Fadin:1998py,Balitsky:2008zza}, the higher order corrections to the impact factors, and the corrections to the single inclusive hadron production~\cite{Chirilli:2012jd,Xiao:2014uba,Watanabe:2016gws}. There has thus been an increased interest in incorporating these higher order calculations into phenomenological studies.

Recently, the forward single inclusive hadron production in proton (or deuteron)-nucleus collisions
has been analyzed up to next-to-leading order accuracy~\cite{Xiao:2014uba} based on the analytical calculation presented in \cite{Chirilli:2012jd}. It was shown that the NLO corrections are large and can overtake the leading order term at higher values of transverse momenta.

In this review, we discuss this calculation as well as possible ways to remedy the negativity problem, including the kinematic corrections that go beyond the high energy limit. The structure of this document is as follows. In the next section we shall briefly recap the idea of parton saturation and the nonlinear evolution equations which aim to include this effect. In Sec.~\ref{sec:forward inclusive LO}, we shall introduce the framework for the calculation of forward hadron production in the high energy limit, which includes rescattering corrections. In Sec.~\ref{sec:forward inclusive NLO}, we will present the outline of the NLO calculation, and the comparison with experimental data. We will also discuss the origin of the negativity and  present different ways to improve the calculation, among them the inclusion of kinematical effects.

\section{Parton saturation and nonlinear evolution equation}
\label{sec:evolution}

At small values of the Bjorken variable $\bjorkenx$, integrated parton densities exhibit a very strong growth as $\bjorkenx$ decreases. This phenomenon is characteristic of a massless non-abelian theory like QCD, and is due to the multiple splitting of gluons. It can be explained qualitatively in the following way. Consider an initial state which contains a single quark or gluon. The subsequent emission of a gluon occurs with the probability
\begin{equation}
 \dd\probability \simeq \frac{C\alphas}{\pi^2} \frac{\dd[2]\kperp}{\kperp^2}\frac{\dd\bjorkenx}{\bjorkenx} \; ,
\label{eq:fac}
\end{equation}
where $\kperp$ is the transverse momentum of the emitted gluon with respect to the initial particle and $\gluonsplitting$ is the fraction of the longitudinal momentum of the initial particle. The color factor $C$ depends on the type of the parent particle. Now, each subsequent emission can proceed from the quark or the gluon, bringing in another copy of the factor associated with Eq.~\eqref{eq:fac}. Let us consider here the situation in the high energy limit in which the transverse momenta of the emitted gluons are comparable but the longitudinal momenta are strongly ordered.  Even if $\alphas$ is small, it can be compensated by the large logarithm $\ln 1/\gluonsplitting$. The subsequent emissions carry ever smaller values of $\gluonsplitting$, i.e. emissions with strong ordering in the longitudinal momentum
\begin{equation}
 \gluonsplitting_n \ll \gluonsplitting_{n-1} \ll \dots \ll \gluonsplitting_2 \ll \gluonsplitting_1 \; \; ,
\end{equation}
will result in terms proportional to $(\alphas \ln 1/\gluonsplitting)^n$.
For small values of $\gluonsplitting$, these are large and need to be resummed. This resummation is accomplished in the Regge limit by the BFKL equation~\cite{Balitsky:1978ic,Kuraev:1977fs}, which is the evolution equation for the unintegrated gluon density derived in the high energy limit. 
It can be cast in the following form
\begin{equation}
 \pdv{f(\gluonsplitting,\kperp)}{\ln 1/\gluonsplitting} = \int \frac{\dd \kperp'^2}{\kperp'^2} \, K(\kperp,\kperp') \, f(\gluonsplitting,\kperp') \; ,
\label{eq:bfkl_LO}
\end{equation}
where $f(\gluonsplitting, \kperp)$ is the unintegrated gluon density, which depends on the gluon transverse momentum $\kperp$. In the small-$\gluonsplitting$ approximation, the unintegrated gluon density $f(\gluonsplitting, \kperp)$ is related to the  more standard integrated gluon density through the following relation
\begin{equation}
 \partondist{g}{\gluonsplitting}{Q^2} = \int^{Q^2} \frac{\dd \kperp^2}{\kperp^2}  \, f(\gluonsplitting, \kperp) \, .
\end{equation}

The function $\bfklkernel(\kperp, \kperp')$ is the evolution kernel of the BFKL equation, which gives the branching probability for the gluons in the small-$\gluonsplitting$ limit, and which has the following expansion in terms of powers of $\alphas$:
\begin{equation}
 \bfklkernel(\kperp, \kperp')  = \alphasbar \bfklkernel_{0}(\kperp, \kperp') + \alphasbar^2 \bfklkernel_{1}(\kperp, \kperp') + \order{\alphasbar^3}\, ,
\label{eq:kernel_alphas}
\end{equation}
where $\alphasbar = \alphas \Nc/\pi$, and $\bfklkernel_0$ and $\bfklkernel_1$ are leading logarithmic (LL) and next-to-leading logarithmic (NLL) kernels, computed respectively in Refs.~\cite{Balitsky:1978ic,Kuraev:1977fs} and~\cite{Camici:1997ij,Ciafaloni:1998gs,Fadin:1998py}. The evolution kernels contain real and virtual parts, which are divergent as $\kperp'\to \kperp$, but when combined together insure the infrared safety of the evolution. The solution to this equation has been constructed in Ref.~\cite{Lipatov:1985uk} (for a recent derivation of the solution in the NLL case see \cite{Chirilli:2013kca}). It can be shown that the solution to the LL BFKL equation behaves like a power with decreasing $\gluonsplitting$, 
\begin{equation}
 f(\gluonsplitting, \kperp) \sim \gluonsplitting^{-\lambda}, \qquad \lambda = 4 \ln 2 \alphasbar \quad (\text{LL result})
\label{eq:pomeron_LO}
\end{equation}
This behavior is known as the hard Pomeron behavior, as opposed to the soft Pomeron, which would have a power $\lambda \sim 0.08$, and which was used in the phenomenology of hadronic collisions.

This result poses two immediate problems. The first one is that such a strong power growth  is not phenomenologically supported, since the result \eqref{eq:pomeron_LO} would give about $\lambda \sim 0.5$ for typical values of the strong coupling , whereas the power extracted from the experimental data, primarily on deep inelastic scattering of leptons off protons, is approximately $0.25-0.3$. This is the power seen in the gluon distribution function extracted from the proton structure function. The second problem is with the unitarity of the scattering amplitude. The gluon density itself and the corresponding cross section can grow without bound; however, the scattering amplitude at fixed impact parameter has to obey a unitarity bound. This poses a constraint on the possible functional form of the growth of the gluon density itself, when the scattering amplitude is integrated over the impact parameter.  The Pomeron solution in the form of the power, as in Eq.~\eqref{eq:pomeron_LO}, thus violates the unitarity bound. 

The NLL corrections to the evolution kernel have been computed in Refs.~\cite{Camici:1997ij,Ciafaloni:1998gs,Fadin:1998py}, and more recently in the context of the dipole evolution in~
\cite{Balitsky:2008zza}. There are several physical sources of the corrections, which we can classify as follows:
\begin{itemize}
\item running coupling corrections
\item corrections from the kinematics
\item corrections from the non-singular (in $1/\gluonsplitting$) terms of the DGLAP splitting function
\end{itemize}

\begin{figure}
 \centering
 \includegraphics[width=8cm]{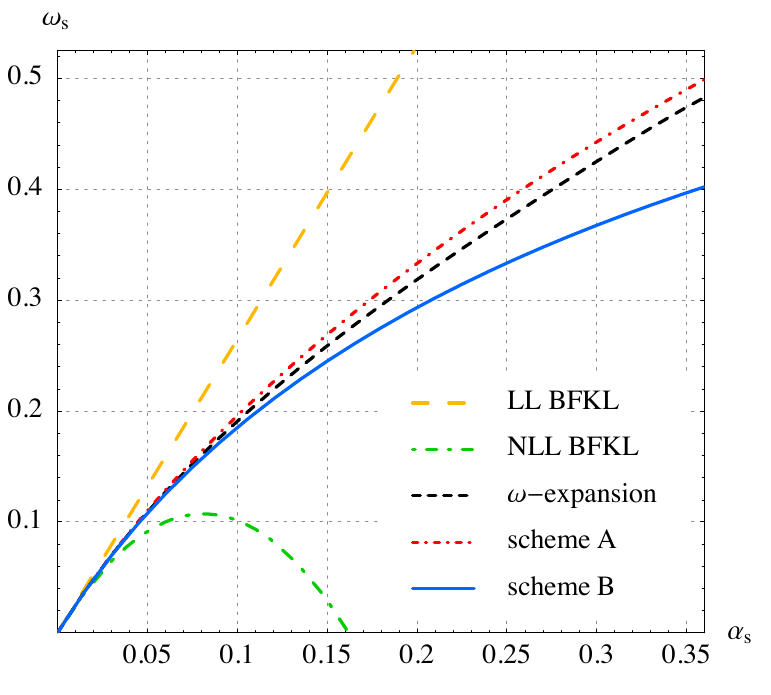}
 \caption{The effective intercept of the BFKL solution in the case of the leading logarithmic (yellow dashed) and next-to-leading logarithmic (green dotted-dashed) approximation. Also shown are different resummation schemes (red dotted - dashed, black dashed and blue solid). Figure reproduced from \cite{Ciafaloni:2003rd}.}
 \label{fig:bfklllnll}
\end{figure}

In LL order, the BFKL kernel has fixed coupling, and in fact its kernel is identical to that in $N=4$ SYM theory at this level, see for example Ref.~\cite{Balitsky:2009xg}. It is only at the NLL level that the coupling starts to run in the small-$\bjorkenx$ formalism in QCD. Clearly, the running of the coupling is one of the most important corrections to be incorporated into phenomenological calculations. The second class of corrections stems from the kinematics. For example, in the LL order, the working assumption is that the rapidity difference between two subsequent gluon emissions is large enough so that the leading logarithm in energy is picked up. At NLL there is a large correction from the kinematical situation when the two gluons are not necessarily very distant in rapidity. This leads to the corrections from the choice of scale at this order, see Ref.~\cite{Ciafaloni:1998gs}. The third class of corrections stems from the non-singular parts of the DGLAP splitting function. In the double logarithmic limit, the BFKL and DGLAP equations actually coincide, with the $1/\gluonsplitting$ part of the gluon splitting function being the dominant one and common to both equations. At the NLL level, the BFKL equation contains subleading (in $1/\gluonsplitting$) terms of the DGLAP splitting function. 
All these corrections are numerically very large, as compared with the LL calculation. In Fig.~\ref{fig:bfklllnll} we show the intercept of the solution to the BFKL equation at LL and NLL  orders. It is evident that the NLL corrections are large, even at very small values of the coupling constant, and can lead to the negative intercept. Also shown are different resummation schemes which stabilize the result.

Apart from the NLO corrections described above, there are also other classes of corrections that arise due to the presence of the high gluon density. It is expected from QCD that when the energy is very high, or correspondingly $\gluonsplitting$ is very small, the gluon distribution described by Eq.~\eqref{eq:bfkl_LO} is so large due to the gluon splitting that the competing mechanism of gluon recombination becomes significant. The high density of gluons leads to the screening of gluons in transverse space, and as a result the growth is tamed by the presence of the additional terms in the evolution equation. These additional terms are  nonlinear in the density and enter the evolution equation with a negative sign, leading to the phenomenon known as parton saturation~\cite{Gribov:1984tu,Mueller:1985wy}. The basic equation that incorporates these effects is the Balitsky-Kovchegov (BK) equation, which was independently derived in Ref.~\cite{Balitsky:1995ub} using the operator product expansion for high energy scattering and in Ref.~\cite{Kovchegov:1999yj} from the Mueller dipole formalism~\cite{Mueller:1993rr} at small $\gluonsplitting$. To be precise, the Balitsky formalism gives an infinite hierarchy of coupled equations for the correlators of Wilson lines. In the large-$\Nc$ limit, the first equation decouples from the rest of the hierarchy and can be solved independently. In this approximation it coincides with the single equation derived in Ref.~\cite{Kovchegov:1999yj}.

The BK equation is an equation for the dipole scattering amplitude $\dipoleN$, which can be related to the dipole unintegrated gluon density in momentum space as follows
\begin{equation}
 \dipoleF(\kperp) = \int \frac{\dd[2]\vec\xperp \dd[2]\vec\yperp}{2\pi^2} e^{-i \vec\kperp \cdot (\vec\xperp - \vec\yperp)} \, \bigl(1 - \dipoleN(\vec\xperp, \vec\yperp)\bigr) \, .
\end{equation}
This is the amplitude for the scattering of a quark-antiquark dipole off a potentially dense target. The gluon branching is then described as the evolution of this amplitude, i.e. the splitting of the dipole into daughter dipoles. The BK equation for the dipole evolution can be cast in the following form
\begin{equation}
 \pdv{\dipoleN(\vec\xperp, \vec\yperp)}{\evolutionrapidity} = \int\frac{\dd[2] \vec\bperp}{2\pi} \bkkernel\bigl[\dipoleN(\vec\xperp, \vec\bperp) + \dipoleN(\vec\bperp, \vec\yperp) - \dipoleN(\vec\xperp, \vec\yperp) - \dipoleN(\vec\xperp, \vec\bperp)\,\dipoleN(\vec\bperp, \vec\yperp)\bigr] \, .
\label{eq:BK}
\end{equation}
The two  vectors  $\vec\xperp$ and $\vec\yperp$ denote the positions of the dipole endpoints in the two-dimensional transverse coordinate space. The branching kernel $\bkkernel$ depends on the dipole sizes involved and contains all information about the splitting of the dipoles. In addition, it also depends on the running coupling $\alphas$.  It can be demonstrated that the linear part of Eq.~\eqref{eq:BK}, when transformed into momentum space, is equivalent to the linear BFKL equation. The additional nonlinear term that appears in Eq.~\eqref{eq:BK} is responsible for the parton saturation and tames the growth of the gluon density.  As is clear from the form of Eq.~\eqref{eq:BK}, the nonlinear term will reduce the growth of the dipole amplitude. In fact $\dipoleN = 1$ is a fixed point of this equation, and the solution will saturate to that value after a sufficiently long rapidity ($\evolutionrapidity = \ln 1/\gluonsplitting$) interval. The BK equation is now known up to the NLL order~\cite{Balitsky:2008zza} and at this higher order its form is more complicated than shown in Eq.~\eqref{eq:BK} (see discussion later in this section).

In the LL order, the branching kernel has the form
\begin{equation}
 \bkkernel(\vec\xperp, \vec\yperp, \vec\bperp; \alphas) = \frac{\alphas\Nc}{\pi} \frac{(\vec\xperp - \vec\yperp)^2}{(\vec\xperp - \vec\bperp)^2(\vec\bperp - \vec\yperp)^2} \; .
\label{eq:BKkernel}
\end{equation}
 Note that the kernel depends only on the differences between the dipole endpoints and not on the absolute coordinate positions (unlike the dipole amplitudes which depend on both the coordinate differences --- dipole sizes ---- and the coordinate sums --- the dipole impact parameter).

\begin{figure}[htbp]
\begin{center}
\includegraphics[width=8cm]{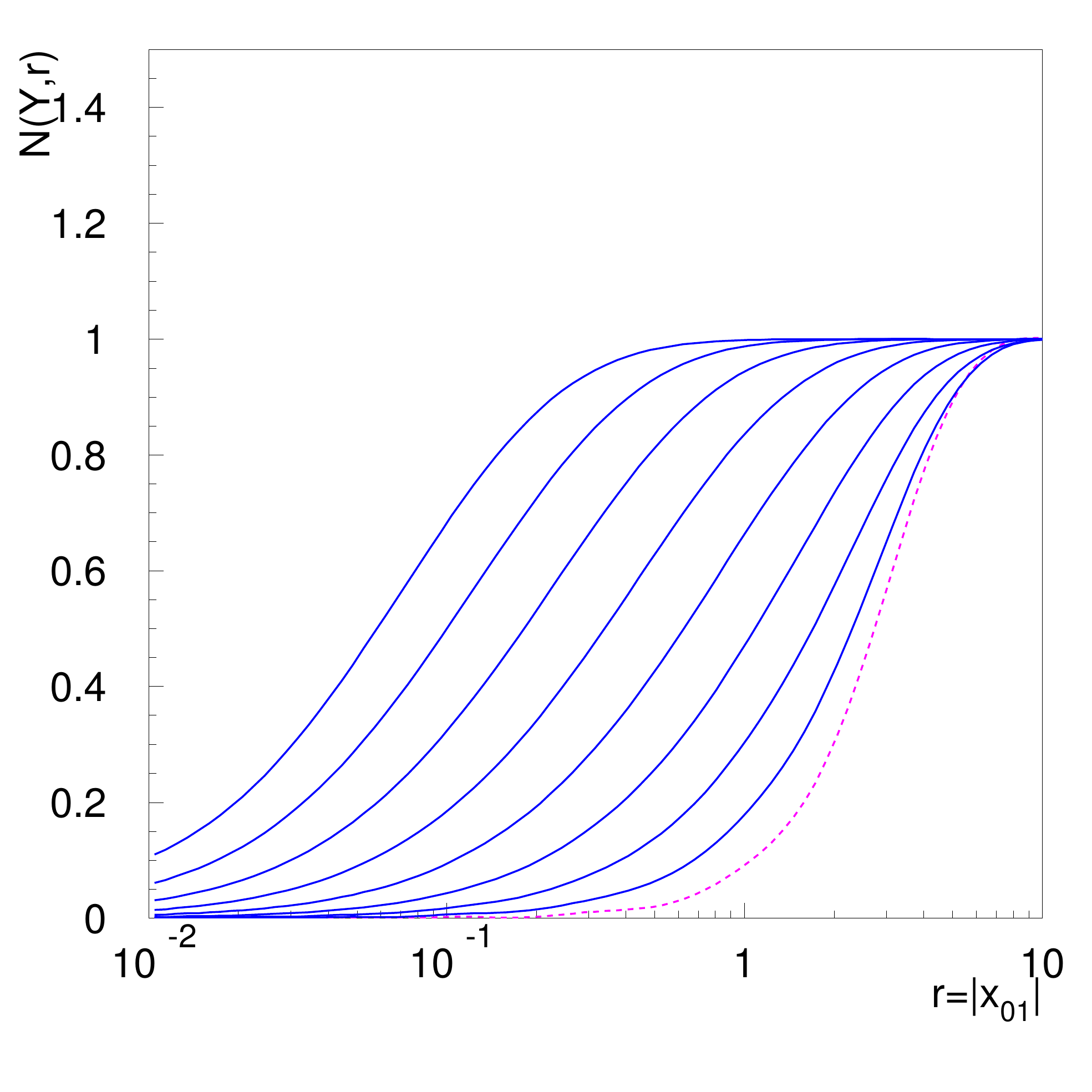}
\caption{The solution to the nonlinear BK equation as a function of the dipole size $r$ for different values of the rapidity. Different curves (solid blue) from right to left denote the different values in rapidity $\evolutionrapidity = 1.5,\dots,15.6$. The dotted magenta line denotes the initial condition for the evolution equation. The horizontal axis is in arbitrary units.}
\label{fig:bksol}
\end{center}
\end{figure}

The BK equation is usually solved under a simplifying assumption that the impact parameter is neglected, i.e. that the dipole amplitude depends only on the dipole size $\vec\rperp = \vec\xperp - \vec\yperp$. The solution for that case is plotted in Fig.~\ref{fig:bksol}, where the dipole amplitude is shown as a function of the dipole size $\rperp$ for a fixed value of rapidity $\evolutionrapidity$. The dipole amplitude is small for small values of the dipole size, i.e. the color transparency phenomenon, and saturated to unity for large values of dipole size. With increasing rapidity $\evolutionrapidity$ the amplitude grows as well, and the point at which it becomes substantial moves to lower values of the dipole sizes. We observe that the solution exhibits a front in $\rperp$ which moves towards smaller values of $\rperp$ as rapidity increases. This can be quantified by introducing the saturation scale $\satscale(\evolutionrapidity)$, which is the characteristic scale at which the amplitude becomes large and the nonlinear effects become important. The saturation scale can be defined by
\begin{equation}
 \dipoleN(\rperp = 1/\satscale, \evolutionrapidity) = \kappa \; ,
\label{eq:satscale}
\end{equation}
where $\kappa$ is some constant number, for example $1/2$.
Eq.~\eqref{eq:satscale} will result in the rapidity dependence of the saturation scale $\satscale(Y)$. The functional dependence of the saturation scale on the rapidity is approximately exponential $\satscale \sim \exp(\lambda_s \evolutionrapidity) \sim \gluonsplitting^{-\lambda_s}$. We stress that the growth of the saturation scale with rapidity can be computed from the evolution equation and thus is derived from perturbative QCD. On the other hand the normalization cannot be extracted from the evolution itself, but it depends on the initial conditions which  include the non-perturbative physics, and also on the type of the target, i.e. proton versus nucleus.

\begin{figure}[htbp]
\begin{center}
\includegraphics[width=8cm]{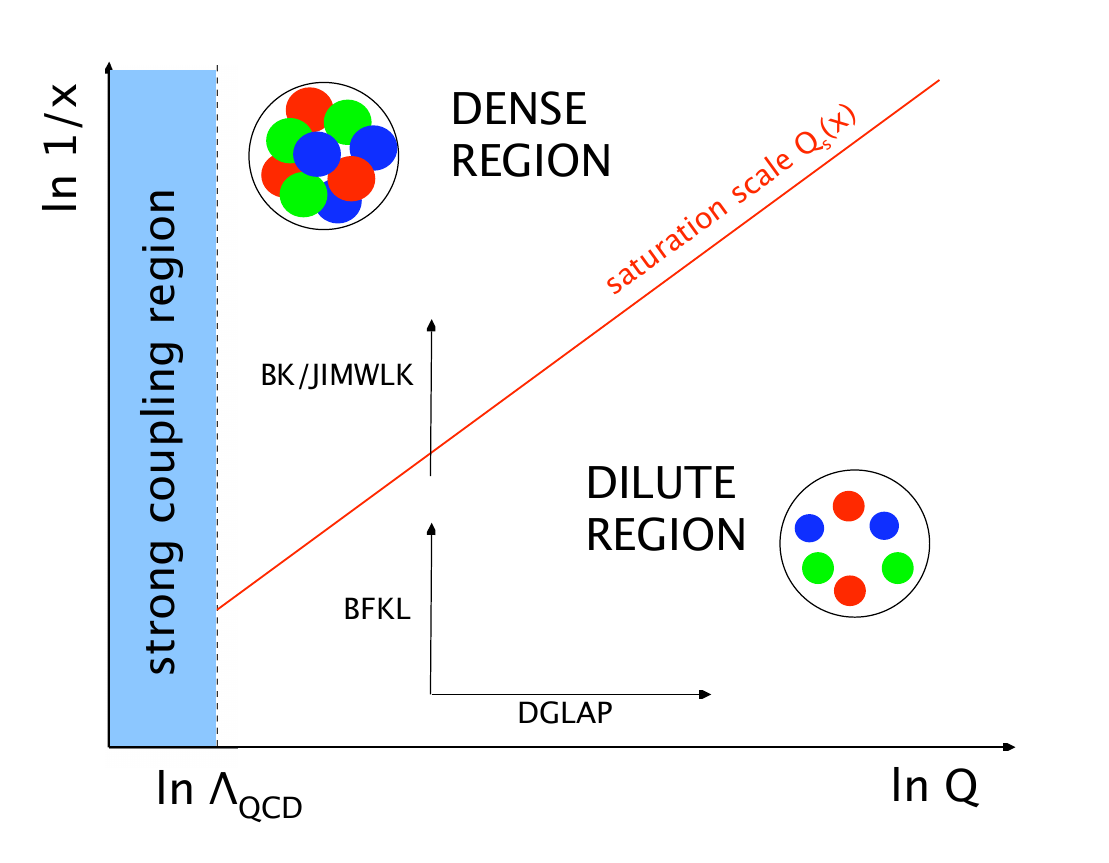}
\caption{Schematic illustration of the different types of evolution in the $(\gluonsplitting, Q)$ plane. The diagonal line is the saturation scale which divides the dilute and dense partonic regime.
The plot is taken from Ref.~\cite{AbelleiraFernandez:2012cc}.}
\label{fig:saturation}
\end{center}
\end{figure}

The saturation scale thus divides two regions of different parton densities: the dilute regime, such that $\rperp < 1/\satscale$, with scales higher than the saturation scale, and the dense regime $\rperp > 1/\satscale$, where the nonlinear effects need to be taken into account. This is illustrated in Fig.~\ref{fig:saturation}. The horizontal axis is related to the momentum scale $Q$, which could be roughly related to the inverse of the transverse coordinate (dipole size, $\rperp \sim 1/Q$). It characterizes the resolution of the process. The vertical axis, 
is given by $\ln 1/\gluonsplitting$, which is more related to the available energy. We see that the saturation scale is denoted by a curve in this $(Q^2, \gluonsplitting)$ plane, and divides the dilute and dense regimes. In the dilute regime, the linear evolution is applicable; either the BFKL evolution which predicts changes along the $\gluonsplitting$ axis, or the more standard DGLAP evolution which predicts changes along the $Q^2$ axis. In the dense regime, nonlinear effects in the density need to be taken into account and nonlinear evolution equations are required.

We stress that the transition in this diagram is not an abrupt one, but rather is smooth, with saturation being defined up to a normalization factor. It is sometimes also useful to introduce the geometrical scaling region, which is the region where the amplitude depends only on the ratio $\satscale \rperp$, rather than $\rperp, \evolutionrapidity$ separately. This regime encloses the deeply saturated regime and also part of the transition regime. 
Geometric scaling is a property of the solution to the nonlinear equation in the leading logarithmic approximation in $\ln 1/\gluonsplitting$. However, it may be violated if higher order corrections are included. In particular, it was shown from the analysis of the DGLAP evolution with running coupling above the saturation scale~\cite{Kwiecinski:2002ep} that the geometrical scaling is indeed violated. The violation, however, can be factored out and its size is controlled by the parameter $\alphasbar(\satscale^2)\ln Q^2/\satscale^2(\gluonsplitting)$. Therefore, in the region where $\gluonsplitting \ll 1$ and $\ln Q^2/\satscale^2 \ll \ln \satscale^2/\LambdaQCD^2$ the geometrical scaling is preserved. Similar conclusions were also reached in Ref.~\cite{Iancu:2002tr}, where this observation was referred to as an extended geometrical scaling window.

If the impact parameter is not neglected in the evolution equation, then the complete solution becomes rather complicated as it starts to depend on five variables: rapidity, dipole size, impact parameter and two angles. At first it would seem that since the kernel~\eqref{eq:BKkernel} does not depend on the impact parameter, that this variable would not play a major role in the evolution. However, the impact parameter comes into the evolution because the dipole amplitudes $\dipoleN$ depend on it. As a matter of fact, the dipole size and the impact parameter are closely interconnected with each other. As shown in Refs.~\cite{GolecBiernat:2003ym} and~\cite{Berger:2010sh}, once the initial condition includes the profile in the impact parameter dependence, the subsequent evolution typically changes the initial form quite rapidly. For example, when the initial profile is exponential in impact parameter, the evolution will modify it into power-like behavior, since the kernel is scale invariant and therefore the interaction is long-range. This is unphysical, because QCD exhibits confinement, and therefore there is a mass gap in the theory which results in the finite range of the strong interaction. Therefore, as it stands at the moment, the BK equation is incomplete as it does not include this vital information, and thus the kernel has to be regulated by the mass parameter. This mass parameter needs to be included essentially by hand. In other words the perturbative BK evolution does indeed preserve the  unitarity condition of the dipole amplitude through the perturbative saturation, but the resulting growth in energy of the integrated cross section (over the impact parameter) will still take the form of a power law~\cite{Kovner:2001bh,Kovner:2002yt}. This will violate the Froissart bound which was derived under the assumption of the finite range of the interaction~\cite{Froissart:1961ux}.

In the following we will not discuss the impact parameter dependence, since the observables that we will study are sufficiently inclusive and not sensitive to the impact parameter profile. Nevertheless, the proper modeling of the impact parameter dependence is essential for many of  the phenomenological applications, in particular for many more exclusive reactions, though not only restricted to those (see Refs.~\cite{Berger:2011ew,Berger:2012wx} for phenomenological studies using full impact parameter dependence). 

Recently, there has been a lot of research activity concerning the nonlinear evolution at NLL order and beyond. The original calculation of the nonlinear evolution at NLL was performed in Ref.~\cite{Balitsky:2008zza}, where the linear limit of this calculation coincided with the linear BFKL evolution at NLL order derived earlier~\cite{Fadin:1998py,Ciafaloni:1998gs}. The numerical analysis of this nonlinear evolution equation at NLL was first performed in Ref.~\cite{Lappi:2015fma}, where it was demonstrated that the NLL corrections are large and lead to instability of the solution. Following this work, a resummation procedure was proposed~\cite{Iancu:2015joa,Iancu:2015vea} to stabilize the solution, based on collinear improvements, which is essentially analogous to the resummation proposed earlier in Refs.~\cite{Andersson:1995ju,Kwiecinski:1996td,Kwiecinski:1997ee,Salam:1998tj,Ciafaloni:2003rd,Ciafaloni:2003ek,Vera:2005jt},  which was applied to the linear case. The solution with the resummation was shown to be numerically stable~\cite{Lappi:2016fmu,Iancu:2015joa,Iancu:2015vea}.

\section{Forward inclusive hadron production at LO}\label{sec:forward inclusive LO}

In the previous section, we have introduced the concept of parton saturation and how can it be described through the nonlinear evolution equations derived in QCD. The major question is whether this phenomenon is present in hadron collisions at currently attained collider energies, and how to observe it in experimental data and  best quantify it.

There have been many phenomenological applications of the small $\bjorkenx$ formalism which include parton saturation effects. Among them are the calculations  of the inclusive structure function at HERA~\cite{GolecBiernat:1998js,GolecBiernat:1999qd,Stasto:2000er,Gotsman:2002yy,Albacete:2009fh}, diffraction and  vector meson production~\cite{Munier:2001nr,Gotsman:2003br,Rogers:2003vi,Kowalski:2006hc,Berger:2012wx,Lappi:2013am}, and also multiplicities at RHIC and LHC in proton-proton and heavy nucleus collisions~\cite{Albacete:2007sm}, to name just a few. In this review we shall focus on the inclusive forward production of single hadrons in proton-nucleus collisions. In this section we shall describe the special formalism for the calculation of this process in the high energy limit and its extension beyond the lowest order of accuracy. 

\begin{figure}
 \centering
 \includegraphics{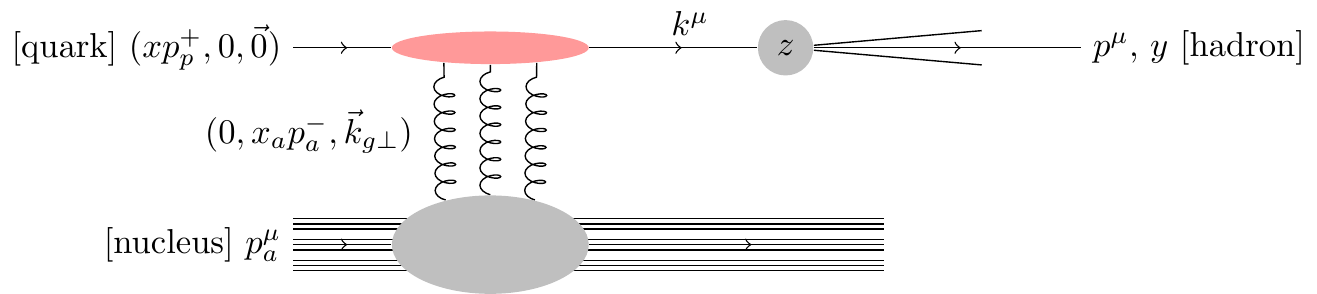}
 \caption{Kinematics of the leading order process for the forward inclusive hadron production. In this illustration the quark from the incoming projectile interacts with the dense gluon field of the nucleus and emerges with the additional transverse momentum. Finally it hadronizes into the hadron which is detected experimentally. Similar process exists for the initial state gluon.}
 \label{fig:lo-kinematics}
\end{figure}

We start the small-$\bjorkenx$ description of the forward production in $\pA$ collisions by considering the scattering of a quark on a nucleus, which is illustrated in Fig.~\ref{fig:lo-kinematics}. Multiple scattering of the quark off the gluons in the field of the nucleus can be encompassed in the Wilson line
\begin{equation}
 \Uwilson(\xperp) = \pathordered \exp\biggl(i \strongcoupling \int_{-\infty}^{+\infty} \dd x^+ \, T^a A_a^{-}(x^+,\xperp)\biggr) \; ,
\label{eq:uwilson}
\end{equation}
where the integral is over the path of the quark which is traveling along the $x^+$ direction, and $A_a^{-}(x^+, \xperp)$ is the gluon field of the nucleus, the solution of the classical Yang-Mills equation. Here, $T^a$ is an $SU(3)$ generator matrix in the fundamental representation, and $\strongcoupling$ is the strong coupling.  We shall be working in the high energy or small $\bjorkenx$ approximation, which assumes a large center-of-mass energy between the incoming quark and the nucleus.  We also neglect the recoil of the nucleus.

The lowest order differential cross section for the process of the inclusive quark production, $\pA\to qX$, is then
\begin{equation}
 \frac{\dd[3]\quarkxsec}{\dd\rapidity \dd[2]\vec\kperp} = \sum_f 
\xprojectile q_f(\xprojectile)  \int \frac{\dd[2]\vec\xperp \dd[2]\vec\yperp}{(2\pi)^2} \; e^{-i\vec\kperp \cdot (\vec\xperp - \vec\yperp)} \frac{1}{\Nc} \expval{\trace \Uwilson(\vec\xperp) \Uwilson^{\dagger}(\vec\yperp)}_{\evolutionrapidity} \; ,
\label{eq:quarklo}
\end{equation}
where $\xprojectile = \frac{\kperp}{\sqs}e^{\quarkrapidity}$ is the fraction of the longitudinal momentum of the proton carried by the incoming quark and $\xtarget = \frac{\kperp}{\sqs} e^{-\quarkrapidity}$ is the fraction of the longitudinal momentum of the nucleus carried by the gluon. The final quark is produced with transverse momentum $\kperp$ and at rapidity $\quarkrapidity$.   The color average $\expval{\cdots}_{\evolutionrapidity}$ is understood to be taken over the color sources of the nucleus, as is usually done in the framework of the Color Glass Condensate. The rapidity $\evolutionrapidity \simeq 1/x_g$ is the difference between the rapidity of the gluon and the rapidity of the nucleus.
Performing the CGC color average over the correlator yields the dipole gluon distribution,
\begin{equation}
 \dipoleS(\vec\xperp, \vec\yperp) \defn \frac{1}{N_c} \expval{\trace \Uwilson(\vec\xperp) \Uwilson^{\dagger}(\yperp)}_{\evolutionrapidity} \, .
\end{equation}
Also, $\partondistLO{q_f}{\xprojectile}$ is the quark distribution in the incoming proton, where label $f$ denotes the specific flavor of the incoming quark.

For a complete description of the hadronic cross section it is necessary to include the gluon-initiated channel. To this aim, one has to define the correlators of the Wilson lines in the adjoint representation
\begin{equation}
 \dipoleSadj(\vec\xperp, \vec\yperp) \defn \frac{1}{\Nc^2 - 1} \expval{\trace \Wwilson(\vec\xperp)\Wwilson^{\dagger}(\vec\yperp)}_{\evolutionrapidity} \; .
\label{eq:wwilson}
\end{equation}

One can introduce the Fourier transform into the momentum space of the spatial correlators
\begin{equation}
 \dipoleF(\kperp) \defn \int \frac{\dd[2]\vec\xperp \dd[2]\vec\yperp}{(2\pi)^2} \, e^{-i \vec\kperp \cdot (\vec\xperp - \vec\yperp)} \dipoleS(\vec\xperp, \vec\yperp) \; ,
\label{eq:fk}
\end{equation}
and 
\begin{equation}
 \dipoleFadj(\kperp) \defn \int \frac{\dd[2]\vec\xperp \dd[2]\vec\yperp}{(2\pi)^2} \, e^{-i \vec\kperp \cdot (\vec\xperp - \vec\yperp) } \dipoleSadj(\xperp, \yperp) \; .
\label{eq:fktilde}
\end{equation}
In the large-$\Nc$ limit, one can rewrite the unintegrated gluon distribution in the adjoint representation using the dipole distributions $\dipoleS$, which use the Wilson lines in the fundamental representation (thanks to the identity between Wilson lines in both representations). One thus arrives at the following simplified expression for the unintegrated gluon distribution in the adjoint representation 
\begin{equation}
 \dipoleFadj(\kperp) = \int \frac{\dd[2]\vec\xperp \dd[2]\vec\yperp}{(2\pi)^2} \, e^{-i \vec\kperp \cdot (\vec\xperp - \vec\yperp) } \dipoleS(\vec\xperp, \vec\yperp) \dipoleS(\vec\yperp, \vec\xperp) \; ,
\end{equation}
which is expressed only through the dipole amplitudes in the fundamental representation.

In order to write the cross section for the production of a hadron one needs to convolute the quark and gluon production cross section with the appropriate fragmentation function. The final expression for the  the cross section for the production of a hadron at forward rapidity in the lowest order in the saturation formalism can be expressed as
\begin{equation}
 \frac{\dd[3]\sigmainclusive}{\dd\hadronrapidity \dd[2]\vec\pperp} = \int_{\tau}^1 \frac{\dd\fragfrac}{\fragfrac^2} \biggl[\sum_f 
  \partondistLO{q_f}{\xprojectile} \dipoleF(\kperp) \fragmentationLO{f}{h}{\fragfrac} + \partondistLO{g}{\xprojectile} \dipoleFadj(\kperp) \fragmentationLO{f}{h}{\fragfrac}
 \biggr] \, .
 \label{eq:sigmapt}
\end{equation}
We can express the other kinematical variables (at the parton level) in terms of the hadron transverse momentum $\pperp$, i.e. $\kperp = \pperp/\fragfrac$, $\xprojectile = \frac{\pperp}{\fragfrac\sqs}e^{\hadronrapidity}$, $\tau = \fragfrac\xprojectile$, and $\xtarget = \frac{\pperp}{\fragfrac\sqs}e^{-\hadronrapidity} $. In the above formula $\fragmentationpurefunc{f}{h}$ and $\fragmentationpurefunc{g}{h}$ are fragmentation functions for the fragmenting quark and gluon into hadron correspondingly.
 
There are several important points about this formula. 
The transverse momentum dependence of the produced hadron is generated exclusively from the transverse momentum dependence of the unintegrated gluon distributions in the nucleus $\dipoleF(\kperp)$ and $\dipoleFadj(\kperp)$.  The lowest order process here is a $2\to 1$ process. This is in contrast with the collinear approach where the hard scattering process is $2\to 2$ (at the lowest order) and the transverse momentum dependence is generated through the hard scattering only. The  formalism here includes two types of distributions: it includes the unintegrated parton distribution from the nucleus side and the collinear parton distribution on the hadron side. As such it is highly asymmetric and only applicable at very high rapidities. Also, formally at this order both the parton distribution and fragmentation functions do not possess any scale dependence. For the phenomenological applications however, the scale-dependent parton distribution $\partondist{q}{\xprojectile}{\facscale^2}$ and fragmentation function $\fragmentation{q}{h}{\fragfrac}{\facscale^2}$ have been commonly used.
Finally, at this lowest order, the correlators $\dipoleS$ are rapidity independent. The rapidity dependence can be incorporated through the BK evolution equation, which enters formally at higher order as we shall see in the next section. 

Formula~\eqref{eq:sigmapt} has been extensively used for phenomenology, in particular for the description of the nuclear ratios 
\begin{equation}
 R_{p(d)A}(\pperp, \hadronrapidity) = \frac{\frac{\dd[2]\sigmainclusive}{\dd\hadronrapidity\dd[2]\pperp}}{\Ncoll\frac{\dd[2]\sigmainclusivepp}{\dd\hadronrapidity\dd[2]\pperp}} \; .
\label{eq:nuclear_ratio}
\end{equation}
where $\Ncoll$ is the number of collisions.

\section{Forward inclusive production at NLO}\label{sec:forward inclusive NLO}

\begin{figure}
 \centering
 \includegraphics{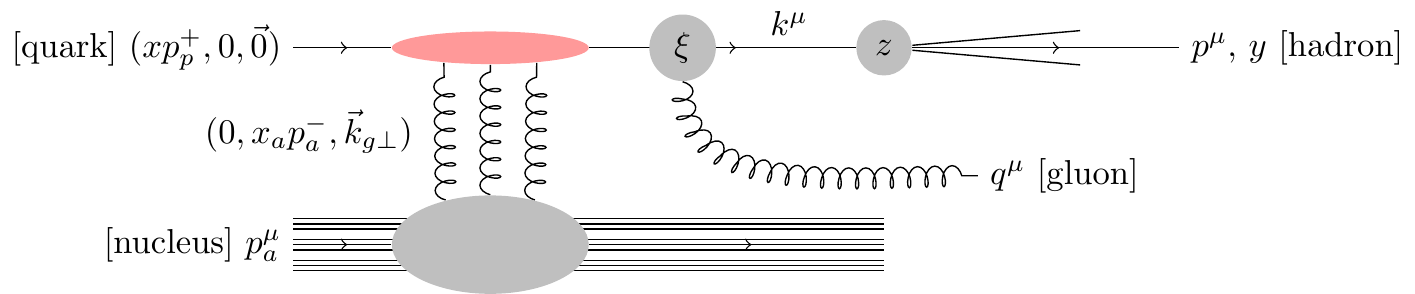}
 \caption{Kinematics of the next-to-leading order process. The initial quark from the incoming projectile undergoes splitting into a quark and gluon, before hadronizing into final state particle. The interaction with the gluon field of the nucleus can occur before the splitting, as shown, or after it.}
 \label{fig:nlo-kinematics}
\end{figure}

Moving beyond leading order, there are two main sources of subleading corrections to the $\pA\to hX$ cross section.
One is the corrections to the BK evolution previously discussed in section~\ref{sec:evolution}.
The next-to-leading corrections to the BK evolution have been computed in Ref.~\cite{Balitsky:2008zza}, and more recently the next-to-leading order form of the more general JIMWLK equation has been obtained in Ref.~\cite{Altinoluk:2014eka}. The dilute limit of the BK equation is the famous BFKL equation. Both calculations, i.e. NLL BK and NLL JIMWLK, reduce to the NLL BFKL~\cite{Fadin:1998py,Ciafaloni:1998gs} in the regime of low density.
The NLL BK is computationally complex, and it has been only solved recently numerically in Refs.~\cite{Lappi:2015fma,Lappi:2016fmu}.
In here we shall focus mostly our attention on the next-to-leading order (NLO) terms in the cross section itself, which result from diagrams in which an unobserved quark or gluon is emitted. The general kinematics of the process is shown in Fig.~\ref{fig:nlo-kinematics}. 
In  this review we will describe the contributions resulting from the one-loop diagrams.

The next-to-leading order calculation of the single inclusive hadron production requires evaluation of several contributions. There are actually four channels to consider. These include two ``diagonal'' channels in which the projectile parton is the same species (quark or gluon) as the one that fragments into a hadron, the quark-quark (qq) channel and the gluon-gluon (gg) channel, as well as two non-diagonal channels, quark-gluon (qg) and gluon-quark (gq), in which the projectile parton and the progenitor of the hadron are different species. Fig.~\ref{fig:nlo-kinematics} shows one of the real diagrams for the quark-quark channel with the kinematics labeled.

\begin{figure}
 \centering
 \includegraphics[width=4.5cm]{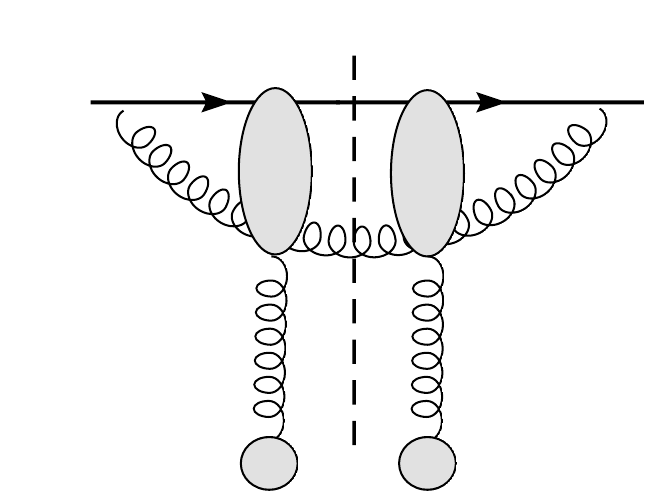}\hspace*{0.5cm}
 \includegraphics[width=4.5cm]{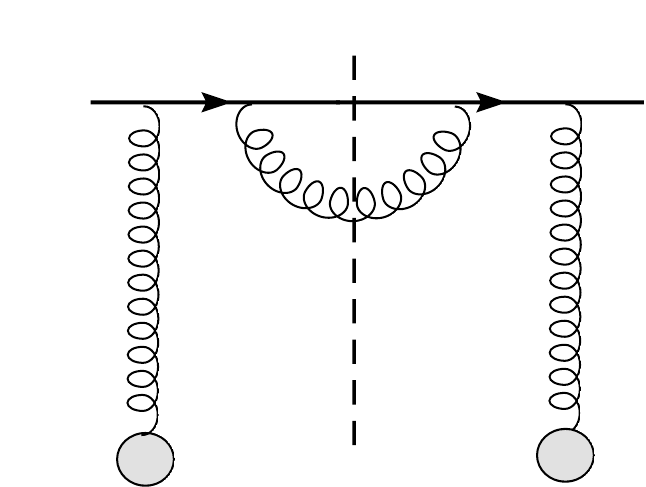}\vspace*{0.5cm}
 \includegraphics[width=4.5cm]{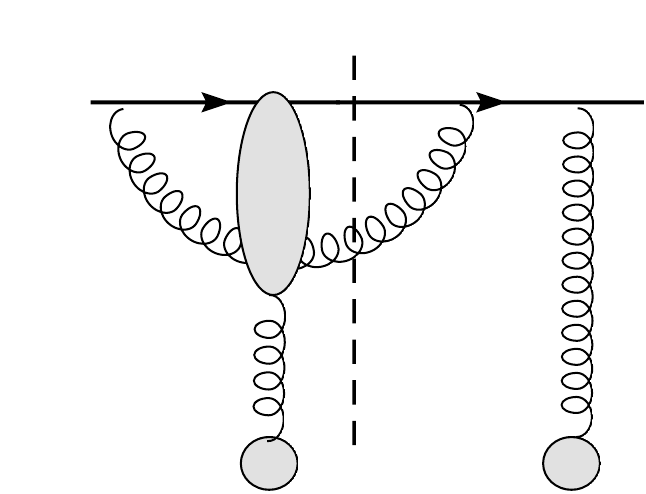}\hspace*{0.5cm}
 \includegraphics[width=4.5cm]{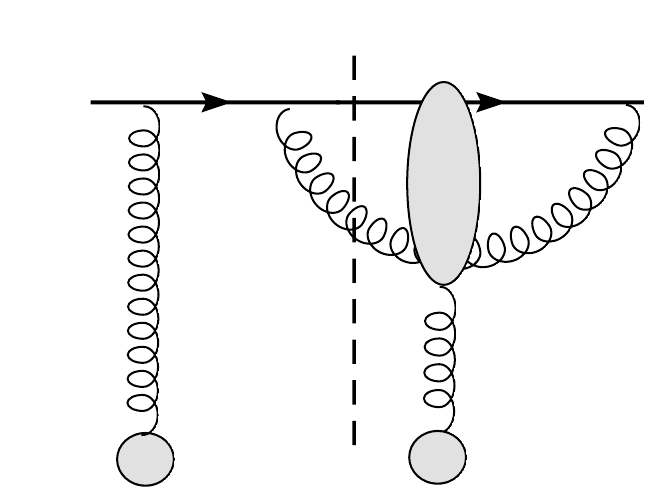}
 \caption{Example of real diagrams for the next-to-leading order quark production $qA\rightarrow qX$. The elliptic blobs denote the interaction of the gluons from the nucleus with the $qg$ system of the initial state. The lower (round) blobs  and the vertical gluons symbolize multiple interactions of these projectile partons with the target nucleus.}
 \label{fig:qtoqreal}
\end{figure}

\begin{figure}
 \centering
\includegraphics[width=4.5cm]{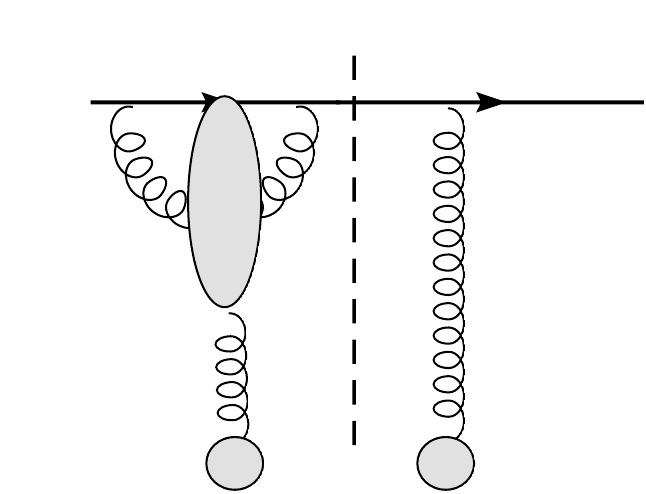}\hspace*{0.5cm}
\includegraphics[width=4.5cm]{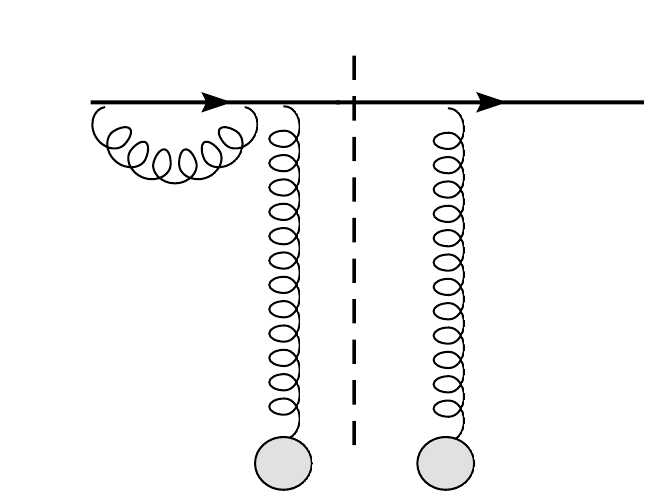}
 \caption{Example of virtual diagrams for the next-to-leading order quark production $qA\rightarrow qX$. The blobs and the vertical gluons symbolize multiple interactions with the target nucleus.}
 \label{fig:qtoqvirtual}
\end{figure}

Several of the real diagrams for the qq channel are shown in Fig.~\ref{fig:qtoqreal}. The quark is the observed particle, and the emitted gluon has to be integrated over. 
In addition to the real diagrams, one needs to include the virtual contributions, as shown in Fig.~\ref{fig:qtoqvirtual}. 

The real and virtual terms need to be combined together to produce the full cross section. These expressions contain different types of divergences, which need to be appropriately subtracted in order to yield finite result. There are rapidity and collinear divergences in the final and initial state, which will be absorbed into the corresponding distributions. Specifically, the rapidity divergence is absorbed into the unintegrated gluon distribution, and the initial and final state collinear divergences are absorbed into the integrated parton distribution and fragmentation functions. The remaining finite contributions are collected in the hard factors.   We shall discuss the  subtractions in more detail in Sec.~\ref{sec:divergences}.

As mentioned above, in addition to the diagonal channel with both real and virtual contributions, there are also non-diagonal channels, where only real contributions exist. Correspondingly, the contribution to the single inclusive cross section from the non-diagonal quark to gluon channel will be obtained by integrating out the emitted quark in the contributions from diagrams in Fig.~\ref{fig:qtoqreal}. 

When all the NLO contributions are included, one arrives at a considerably more complicated expression as compared to the leading order formula, which has the form
\begin{multline}
 \frac{\dd[2]\sigmainclusive}{\dd\hadronrapidity\dd[2]\pperp} = \int_{\tau}^1 \frac{\dd\fragfrac}{\fragfrac^2} \int_{\tau/\fragfrac}^1 \dd\emitfrac
 \Biggl[
 \sum_f \partondist{q_f}{\xprojectile}{\facscale^2} S_{qq} \fragmentation{f}{h}{\fragfrac}{\facscale^2} \\
 + \sum_f \partondist{q_f}{\xprojectile}{\facscale^2} S_{qg} \fragmentation{g}{h}{\fragfrac}{\facscale^2}
 + \sum_f \partondist{g}{\xprojectile}{\facscale^2} S_{gq} \fragmentation{f}{h}{\fragfrac}{\facscale^2} \\
 + \partondist{g}{\xprojectile}{\facscale^2} S_{gg} \fragmentation{g}{h}{\fragfrac}{\facscale^2}
 \Biggr] \; .
 \label{eq:sigmaptnlo}
\end{multline}
There are two key differences that are clear already in this (still simplified) expression, as compared to the leading order result~\eqref{eq:sigmapt}.
First, the unobserved particle, which is integrated over, introduces an additional kinematic degree of freedom. It is  parametrized by $\emitfrac$, the fraction of plus-component momentum retained by the final-state particle which fragments into the hadron.
For the process shown in figure~\ref{fig:nlo-kinematics}, $\emitfrac = \frac{k^+}{k^+ + q^+} = \frac{k^+}{\xprojectile p_p^+}$.
Also, the NLO equation includes ``off-diagonal'' channels, the terms in the second line, where the initial-state particle is a quark and the particle that fragments is a gluon, or vice-versa.
These off-diagonal channels do not exist at leading order, because without the emission of the undetected particle, the initial-state and final-state particles are necessarily of the same species.

The full complexity of the one-loop corrections lies in the expressions for $S_{ij}$, which incorporate both the Wilson line correlators representing the interaction with the target, and the perturbative hard factors representing the emission of the unobserved particle.
The one-loop contributions have been derived and investigated only in the past few years.
In 2011, building on the leading order result~\cite{Dumitru:2005gt,Albacete:2010bs}, Altinoluk and Kovner~\cite{Altinoluk:2011qy} began the investigation of the NLO cross section by incorporating the ``inelastic terms'' which result from projectile partons with high transverse momentum.
They found the following result for the multiplicity, translated from the notation of Ref.~\cite{Altinoluk:2011qy}:
\begin{multline}
 \frac{\dd[3] N_h}{\dd\hadronrapidity\dd[2]\vec\pperp} = \frac{1}{(2\pi)^2} \int_{x_F}^1 \frac{\dd\fragfrac}{\fragfrac^2} \biggl[\partondist{g}{\xprojectile}{\facscale^2} \dipoleFadj(\kperp)  \fragmentation{g}{h}{\fragfrac}{\facscale^2} \\
 \shoveright{+ \sum_f \partondist{q_f}{\xprojectile}{\facscale^2} \dipoleF(\kperp) \fragmentation{q}{h}{\fragfrac}{\facscale^2}\biggr]\qquad} \\
 + \int_{x_F}^1 \frac{\dd\fragfrac}{\fragfrac^2} \frac{\alphas}{(2\pi)^2} \frac{\fragfrac^4}{\pperp^2} \int \frac{\dd[2]\vec\kperp}{(2\pi)^2}\kperp^2 \dipoleFadj(\kperp) \xprojectile \int_{\xprojectile}^1 \frac{\dd\emitfrac}{\emitfrac}\sum_j
  w_{i/j}(\emitfrac) P_{ij}(\emitfrac) f_{j}\biggl(\frac{\xprojectile}{\emitfrac}, \facscale^2\biggr) \fragmentation{i}{h}{\fragfrac}{\facscale^2}\; .
\end{multline}
Here $P_{ij}$ are splitting functions and $w_{i/j}$ are the inelastic weight functions defined in Ref.~\cite{Altinoluk:2011qy}.
The first two lines in the above equation are the elastic LO terms, the same as before in Eq.~\eqref{eq:sigmapt}.
The inelastic terms are in the third line.
These terms were  derived under the assumption that the projectile partons enter with low transverse momentum, and acquire large transverse momentum from hard collisions with gluons in the target nucleus.
The elastic terms are accounted for by the leading order formula~\eqref{eq:sigmapt}, and the inelastic terms represent the simplest NLO contribution.

The first numerical calculation to include these results was presented in Ref.~\cite{Albacete:2012xq}. 
Already the inelastic terms displayed several interesting features.
They found that the terms are negative and have a steeper dependence on transverse momentum than the elastic terms.
In fact, the inelastic contributions completely overwhelm the elastic contributions above some cutoff momentum, which depends on the model chosen for the unintegrated gluon distribution.
This could be expected because the inelastic terms are roughly proportional to $\ln(\pperp/\satscale)$ while the elastic terms are roughly proportional to $-\ln(\pperp/\LambdaQCD)$~\cite{Altinoluk:2011qy,Albacete:2012xq}, making the ratio
\begin{equation}
 r = \frac{\text{elastic}+\text{inelastic}}{\text{elastic}} \sim \frac{\ln(\satscale/\LambdaQCD)}{\ln(\pperp/\LambdaQCD)} \; ,
\end{equation}
This form of the expression helps justify the observation that $r$ drops with increasing hadron transverse momementum $\pperp$.

Later on, the complete corrections to the cross section  up to the one-loop order were derived in Ref.~\cite{Chirilli:2012jd}.
In this calculation, the expressions $S_{ij}$ correspond to a total of eleven terms, each a convolution of a hard factor and a multipole gluon distribution (a correlator of Wilson lines).
\begin{subequations}
\begin{align}
 S_{qq}
 &= \int \frac{\dd[2] \vec\xperp\dd[2] \vec\yperp}{(2\pi)^2}\dipoleS(\vec\xperp, \vec\yperp) \Bigl[\h02qq + \frac{\alphas}{2\pi}\h12qq\Bigr] \notag\\
 &+ \int \frac{\dd[2]\vec\xperp\dd[2]\vec\yperp\dd[2]\vec\bperp}{(2\pi)^4}\quadrupoleS(\vec\xperp, \vec\bperp, \vec\yperp)\frac{\alphas}{2\pi}\h14qq \\
 S_{qg}
 &= \frac{\alphas}{2\pi}\int \frac{\dd[2]\vec\xperp\dd[2]\vec\yperp}{(2\pi)^2} \dipoleS(\vec\xperp, \vec\yperp)\Bigl[\h{1,1}2qg + \dipoleS(\vec\yperp, \vec\xperp)\h{1,2}2qg\Bigr] \notag\\
 &+ \frac{\alphas}{2\pi}\int \frac{\dd[2]\vec\xperp\dd[2]\vec\yperp\dd[2]\vec\bperp}{(2\pi)^4} \quadrupoleS(\vec\xperp, \vec\bperp, \vec\yperp)\h14qg \\
 S_{gq}
 &= \frac{\alphas}{2\pi}\int \frac{\dd[2]\vec\xperp\dd[2]\vec\yperp}{(2\pi)^2} \dipoleS(\vec\xperp, \vec\yperp)\Bigl[\h{1,1}2gq + \dipoleS(\vec\yperp, \vec\xperp)\h{1,2}2gq\Bigr] \notag\\
 &+ \frac{\alphas}{2\pi}\int \frac{\dd[2]\vec\xperp\dd[2]\vec\yperp\dd[2]\vec\bperp}{(2\pi)^4} \quadrupoleS(\vec\xperp, \vec\bperp, \vec\yperp)\h14gq \\
 S_{gg}
 &= \int \frac{\dd[2] \vec\xperp\dd[2] \vec\yperp}{(2\pi)^2}\dipoleS(\vec\xperp, \vec\yperp) \dipoleS(\vec\yperp, \vec\xperp) \Bigl[\h02gg + \frac{\alphas}{2\pi}\h12gg\Bigr] \notag\\
 &+ \int \frac{\dd[2]\vec\xperp\dd[2]\vec\yperp\dd[2]\vec\bperp}{(2\pi)^4}\dipoleS(\vec\xperp, \vec\bperp)\dipoleS(\vec\bperp, \vec\yperp)\frac{\alphas}{2\pi}\h12q{\bar q} \notag\\
 &+ \int \frac{\dd[2]\vec\xperp\dd[2]\vec\yperp\dd[2]\vec\bperp}{(2\pi)^4}\dipoleS(\vec\xperp, \vec\bperp)\dipoleS(\vec\bperp, \vec\yperp)\dipoleS(\vec\yperp, \vec\xperp)\frac{\alphas}{2\pi}\h16gg
\end{align}
\label{eq:allhardfactors}
\end{subequations}
In the above formulae there is another correlator which is defined as
\begin{equation}
\quadrupoleS(\vec\xperp, \vec\bperp, \vec\yperp) =
 \frac{1}{\Nc^2} \expval{\trace[\Uwilson(\vec\xperp)\Uwilson^{\dagger}(\vec\bperp)]\trace[\Uwilson(\vec\bperp)\Uwilson^{\dagger}(\vec\yperp)]}_{\evolutionrapidity} \; .
\end{equation}
The leading-order hard factors are proportional to the delta function, $\h02qq = \h02gg = e^{-i\vec\kperp\cdot\vec\rperp}\delta(1 - \emitfrac)$, as required to reproduce equation~\eqref{eq:sigmapt}, but the remaining hard factors are considerably more complicated.

In this paper, we don't reproduce the full definitions of the hard factors, instead referring interested readers to the original paper~\cite{Chirilli:2012jd} for the expressions.
We will, however, highlight some of their important features as a guide to the complexity of the next-to-leading order calculation.

It is important to note that the expressions~\eqref{eq:sigmaptnlo} together with Eqs.~\eqref{eq:allhardfactors} exhibit a factorized form with appropriate divergences
being factored out into the corresponding distribution functions. It is by far a non-trivial feature as the divergences which arise in this calculation
are of different origin. We shall discuss these divergences and their subtractions in some detail below.

\subsection{Divergences}\label{sec:divergences}

Most importantly, unlike the leading-order result, the next-to-leading order terms contain divergences, which need to be properly regulated to obtain the finite results represented by equations~\eqref{eq:allhardfactors}.
There are two classes of divergences: rapidity and collinear ones.

\subsubsection{Rapidity divergences}\label{sec:rapidity divergences}

Rapidity divergences arise from integrals of the form 
\begin{equation}
 \int^1\frac{\dd\emitfrac}{1 - \emitfrac}\times\text{finite}(\emitfrac) \; ,
\end{equation}
specifically from the upper endpoint $\emitfrac\to 1$. Above, $\text{finite}(\emitfrac)$ denotes some function which is finite when $\emitfrac\to 1$.
The condition $\emitfrac\to 1$ is kinematically equivalent to the gluon rapidity $\gluonrapidity = \ln\frac{1}{\emitfrac}$ going to $\infty$.
The physical interpretation of this divergence is such  that of  the projectile parton emitting a gluon with large longitudinal momentum in the \emph{opposite} direction.
This gluon is actually indistinguishable from a gluon in the target nucleus, therefore it makes sense to absorb this singularity into the target gluon distribution.
This divergence vanishes if we integrate over the transverse momentum $\kperp$.

 Let's consider the expression for the quark-quark channel only, as an example of how rapidity divergences are removed.
 Prior to the subtractions, the quark-quark channel cross section can be expressed as
\begin{multline}\label{eq:unsubtracted qq cross section}
 \frac{\dd[3]\sigmainclusive}{\dd[2]\pperp \dd\rapidity} =
 \int_{\tau}^1 \frac{\dd\fragfrac}{\fragfrac^2} \xprojectile q(\xprojectile, \facscale)D_{h/q}(\fragfrac, \facscale) \dipoleFbare(\kperp) \\
 +
 \frac{\alphas}{2\pi^2}\int_{\tau}^1 \frac{\dd\fragfrac}{\fragfrac^2} \int_{\tau/\fragfrac}^1 \frac{\dd\emitfrac}{1 - \emitfrac}
 \frac{\xprojectile}{\emitfrac} q\biggl(\frac{\xprojectile}{\emitfrac}, \facscale\biggr)D_{h/q}(\fragfrac, \facscale)
 S_{qq}^{\text{real}} \\
 -
 \frac{\alphas}{2\pi^2}\int_{\tau}^1 \frac{\dd\fragfrac}{\fragfrac^2} \int_{0}^1 \frac{\dd\emitfrac}{1 - \emitfrac}
 \xprojectile q(\xprojectile, \facscale)D_{h/q}(\fragfrac, \facscale)
 S_{qq}^{\text{virt}}\; ,
\end{multline}
 where $\dipoleFbare$ is the unrenormalized dipole gluon distribution as defined in equation~\eqref{eq:fk}, and $S_{qq}^{\text{real}}$, $S_{qq}^{\text{virt}}$ represent the real and virtual NLO terms (which of course depend on $\dipoleFbare$).
 We use $\dipoleFbare$ instead of the conventional $\dipoleF[]^{(0)}$ to denote the bare distribution for consistency with $\dipoleSbare$.
 In the notation of Ref.~\cite{Ducloue:2016shw},\footnote{This is easily translated to the notation of Ref.~\cite{Chirilli:2012jd} by comparing Eqs.~(2)--(5) of Ref.~\cite{Ducloue:2016shw} to Eqs.~(16) and~(20) of Ref.~\cite{Chirilli:2012jd}.}
 \begin{align}
  S_{qq}^{\text{real}} &= (1 + \emitfrac^2)\biggl[\CF\mathcal{I}(\vec\kperp, \emitfrac) + \frac{\Nc}{2}\mathcal{J}(\vec\kperp, \emitfrac)\biggr]\;, \\
  S_{qq}^{\text{virt}} &= (1 + \emitfrac^2)\biggl[\CF\mathcal{I}_v(\vec\kperp, \emitfrac) + \frac{\Nc}{2}\mathcal{J}_v(\vec\kperp, \emitfrac)\biggr]\; .
 \end{align}
 To regulate the divergence, we rewrite the $\emitfrac$ integral in the NLO terms as
 \begin{equation}\label{eq:separate subtraction term}
  \int^1 \frac{\dd\emitfrac}{1 - \emitfrac}f(\emitfrac) = \underbrace{\int^1 \frac{\dd\emitfrac}{(1 - \emitfrac)_+}f(\emitfrac)}_{\text{finite term}} + \underbrace{\int_0^1 \frac{\dd\emitfrac}{1 - \emitfrac}f(1)}_{\text{subtraction term}}\; ,
 \end{equation}
 which follows directly from the definition of the plus prescription.
 We then define the renormalized gluon distribution $\dipoleF[]$ to be the sum of $\dipoleFbare$ and the subtraction terms from the real and virtual NLO contributions.%
 \begin{equation}\label{eq:rapidity subtraction}
  \dipoleF[](\qperp) \defn \dipoleFbare(\qperp) + \frac{\alphas}{2\pi^2}\int_0^1\frac{\dd\emitfrac}{1 - \emitfrac}\Bigl[S_{qq}^{\text{real}} - S_{qq}^{\text{virt}}\Bigr]_{\emitfrac=1}\; .
 \end{equation}
 After combining the bare dipole gluon distribution with the subtraction terms, we can write the cross section entirely in terms of finite expressions: $\dipoleF[]$ and plus-regulated NLO contributions.
\begin{multline}
 \frac{\dd[3]\sigmainclusive}{\dd[2]\pperp \dd\rapidity} =
 \int_{\tau}^1 \frac{\dd\fragfrac}{\fragfrac^2} \xprojectile q(\xprojectile, \facscale)D_{h/q}(\fragfrac, \facscale) \dipoleF(\kperp) \\
 +
 \frac{\alphas}{2\pi^2}\int_{\tau}^1 \frac{\dd\fragfrac}{\fragfrac^2} \int_{\tau/\fragfrac}^1 \frac{\dd\emitfrac}{(1 - \emitfrac)_+}
 \frac{\xprojectile}{\emitfrac} q\biggl(\frac{\xprojectile}{\emitfrac}, \facscale\biggr)D_{h/q}(\fragfrac, \facscale)
 S_{qq}^{\text{real}} \\
 -
 \frac{\alphas}{2\pi^2}\int_{\tau}^1 \frac{\dd\fragfrac}{\fragfrac^2} \int_{0}^1 \frac{\dd\emitfrac}{(1 - \emitfrac)_+}
 \xprojectile q(\xprojectile, \facscale)D_{h/q}(\fragfrac, \facscale)
 S_{qq}^{\text{virt}}
\end{multline}
The NLO terms $S_{qq}^{\text{real}}$ and $S_{qq}^{\text{virt}}$ also depend on $\dipoleFbare$, but the difference between $\dipoleFbare$ and $\dipoleF[]$ is one order of $\alphas$ higher --- in this case, that means $\order{\alphas^2}$, which we assume to be negligible in this calculation.
So we can freely replace $\dipoleFbare\to\dipoleF[]$ within the NLO terms without making any additional changes.
 
One can take the definition of the renormalized gluon distribution~\eqref{eq:rapidity subtraction}, plug in the full expressions for $S_{qq}^{\text{real}}$ and $S_{qq}^{\text{virt}}$, and transform to coordinate space.
The equation becomes \cite{Chirilli:2012jd,Kang:2014lha,Ducloue:2016shw}
\begin{multline}\label{eq:position space subtraction}
 \dipoleS[](\vec\xperp, \vec\yperp) = \dipoleSbare(\vec\xperp, \vec\yperp) \\
  - \frac{\alphas\Nc}{2\pi^2}\int_0^1\frac{\dd\emitfrac}{1 - \emitfrac}\int\dd[2]\vec\bperp\frac{(\vec\xperp - \vec\yperp)^2}{(\vec\xperp - \vec\bperp)^2(\vec\yperp - \vec\bperp)^2}\Bigl[\dipoleS[](\vec\xperp, \vec\yperp) - \quadrupoleS[](\vec\xperp, \vec\bperp, \vec\yperp)\Bigr]
\end{multline}
which looks very similar to the integral form of the BK evolution equation.
However, the similarity is deceptive since there is no evolution in this expression.

Ref.~\cite{Chirilli:2012jd} offers two procedures for artifically introducing the rapidity evolution as required to obtain the BK equation.
One can either shift the upper limit of the integral to $1 - e^{-\rapidity}$, where $\rapidity$ is the rapidity difference between the projectile proton and the target nucleus, or shift the denominator of the integration to $1 - \emitfrac + e^{-\rapidity}$.
Either way, this corresponds to dropping the approximation that the projectile and target are moving at speed $c$, taking them off the light cone.
Taking the derivative with respect to $\rapidity$ then yields the BK equation.
However, introducing the rapidity gap between the proton and nucleus as the evolution variable is somewhat unsatisfying, because the BK equation governs the evolution of a gluon field, a parton-level construct, which should not be sensitive to hadron-level kinematics like the rapidity gap ~\cite{Xiao:2014uba}.

To resolve this issue, we need to carefully consider the physical significance of Eqs.~\eqref{eq:rapidity subtraction} and~\eqref{eq:position space subtraction}.
In the NLO kinematics, the projectile parton undergoes two types of gluon interactions: the scattering off the dense gluon field of the target nucleus, represented by $\dipoleFbare$ or $\dipoleSbare$, and the initial or final state emission, represented by the NLO terms in Eq.~\eqref{eq:unsubtracted qq cross section}.
We might naively consider these two processes to be  separated, as the NLO emission involves a large momentum transfer, while the interaction with the gluon field involves small momentum transfer, making them easily distinguishable.
But in fact, the emitted gluon can carry any momentum allowed by kinematics.
Gluons emitted with very small $q^+$ and small $\qperp$ are actually collinear with the target nucleus, and kinematically indistinguishable from the gluon field of the nucleus itself.
So it makes sense to take the part of the NLO term corresponding to ``slow'' and soft (small $q^+$, small $\qperp$) gluon emission, separate it from the ``fast'' emissions with large $q^+$ (Ref.~\cite{Balitsky:1998kc} justifies this split), and \emph{reinterpret} the emission as scattering off an external gluon field.
This external field should be considered part of the target.
 
This procedure introduces a scale separating the fast and slow gluon fields; basically, a cutoff on how much of the phase space for gluon emissions we are going to absorb into the renormalized gluon distribution.
The cutoff is going to enter Eq.~\eqref{eq:rapidity subtraction} or~\eqref{eq:position space subtraction} through the limits of the $\emitfrac$~integral.
We will return to this issue in Section~\ref{sec:rapidity subtraction}.
 
\subsubsection{Collinear divergences}

Collinear divergences arise from integrals of the form $\int\frac{\dd[2]\kperp'}{(\vec\kperp - \vec\kperp')^2}$ or similar, which after angular integration scales as $1/\kperp'$ as $\vec\kperp'\to\vec\kperp$.
Physically, these contributions correspond to quarks or gluons emitted either in the initial state with momentum parallel to that of the incoming parton, or in the final state with momentum parallel to the outgoing parton.
These divergences should therefore be absorbed by, respectively, the parton distribution function or the fragmentation function.

When regulating the collinear divergences, one has to keep consistency with other parts of the calculation, namely the parton distributions and fragmentation functions.
The commonly used fits for these functions come from expressions derived using dimensional regularization in the $\msbar$ scheme, so the regularization must be done in the same scheme.

For example, applying the $\msbar$ scheme to perform collinear subtractions one can remove the collinear divergences in the quark to quark channel by redefining the quark distribution and fragmentation as follows
\begin{align}
 q(\xprojectile, \facscale) &= q^{(0)}(\xprojectile) - \frac{1}{\hat{\epsilon}} \frac{\alphas(\facscale)}{2\pi} \int_{\xprojectile}^1 \frac{\dd\emitfrac}{\emitfrac} \CF P_{qq}(\emitfrac) q\biggl(\frac{\xprojectile}{\emitfrac}\biggr)\; , \nonumber \\
 D_{h/q}(\fragfrac,\facscale) & = D_{h/q}^{(0)}(\fragfrac)-\frac{1}{\hat{\epsilon}} \frac{\alpha_s(\facscale)}{2\pi} \int_x^1 \frac{\dd\emitfrac}{\emitfrac} \CF P_{qq}(\emitfrac) D_{h/q}\biggl(\frac{\fragfrac}{\emitfrac}\biggr) \; ,
\end{align}
where $1/\hat{\epsilon}=1/\epsilon-\gamma_E+\ln 4\pi$ and $\epsilon$ is the parameter of dimensional regularization ($D=4-2\epsilon$).
Above, $P_{qq}$ is the DGLAP leading order splitting function
\begin{equation}
P_{qq}(\emitfrac) = \frac{1+\emitfrac^2}{(1-\emitfrac)}_+ + \frac{3}{2}\delta(1-\emitfrac) \; .
\end{equation}
In that manner, parton distributions and fragmentation functions obtain the scale dependence in this order of calculation.
Note that the resulting scale and rapidity dependence which shows up at NLO calculation is formally due to the leading order logs: leading in $\ln 1/\xtarget$ and leading in $\ln \facscale$. This is because the leading order calculation of the inclusive production is free from any singularities.
\begin{figure}
 \includegraphics[width=.48\linewidth]{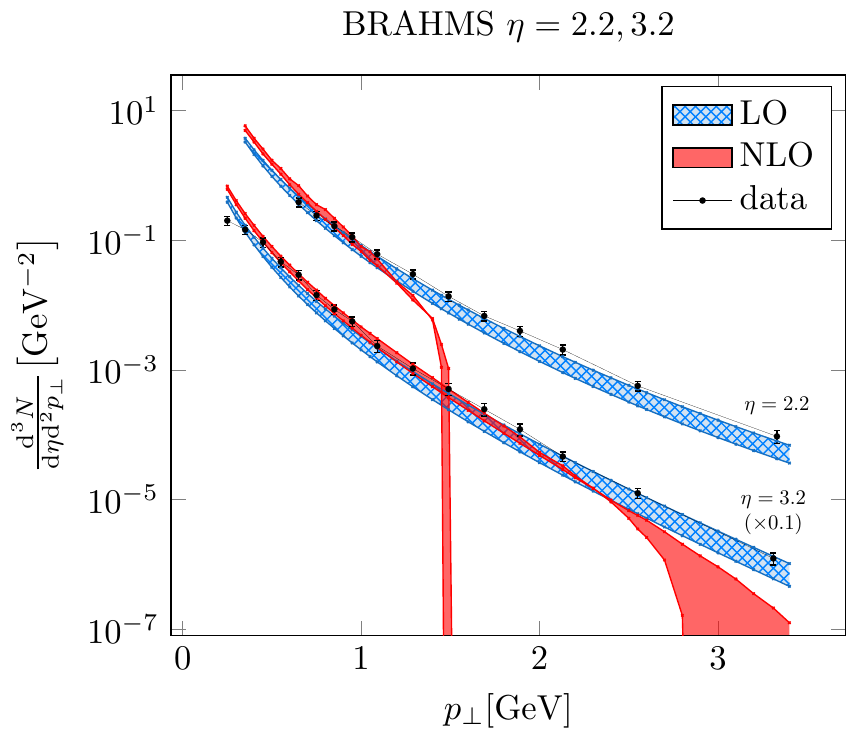}
 \includegraphics[width=.48\linewidth]{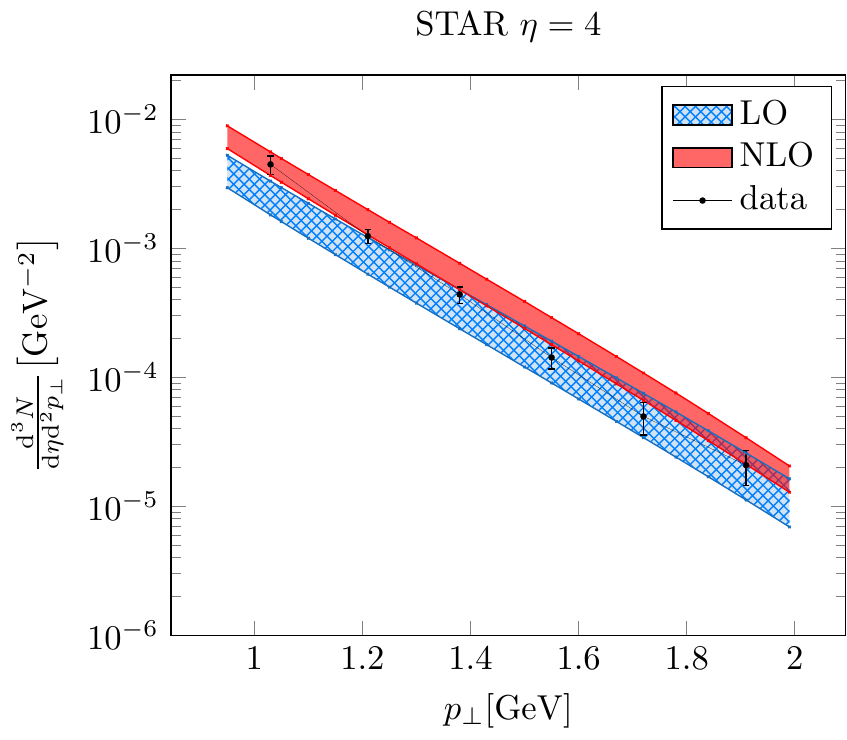}
 \caption{Results of the full NLO cross section, from Ref.~\cite{Stasto:2013cha}.}
 \label{fig:completenloresults}
\end{figure}

\subsection{Numerical results at NLO}

Numerical results from this calculation, using the MSTW 2008 NLO parton distributions\cite{Martin:2009iq} and DSS NLO fragmentation functions\cite{deFlorian:2007hc,deFlorian:2007aj}, were first presented in 2013 by Sta\'sto et al.~\cite{Stasto:2013cha}
The results, shown in Fig.~\ref{fig:completenloresults},  are compared with the experimental data on deuteron-gold collisions measured by BRAHMS~\cite{Arsene:2004ux}
and by STAR~\cite{Adams:2006uz}. Three different models for the unintegrated gluon distribution $\dipoleF(\kperp)$ were used: two phenomenological models, the McLerran-Venugopalan model~\cite{McLerran:1993ni} and the Golec-Biernat-Wusthoff~\cite{GolecBiernat:1998js} model as well as the direct solution to the  leading order BK  equation with the running coupling. 
In general, the agreement between the data and the calculation is very good for low transverse momenta, up to the values of about $\pperp\sim \satscale$. 
The NLO result does  match the experimental data fairly well at lower $\pperp$, down to $\pperp \sim \SI{0.5}{GeV}$ where nonperturbative QCD effects start to dominate.
Within the region of validity, the NLO correction terms do reduce the theoretical uncertainty resulting from the factorization scale and renormalization scale.

\begin{figure}
\begin{center}
 \includegraphics[width=0.6\textwidth]{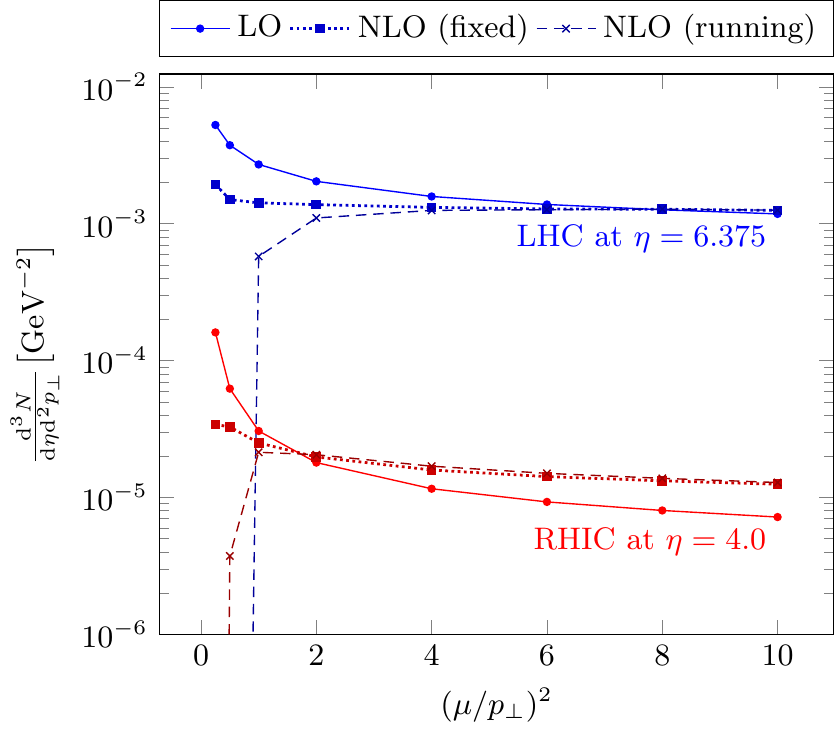}
 \caption{Results of the full NLO calculation, from Ref.~\cite{Stasto:2013cha} showing the scale dependence of the LO (solid lines) and NLO calculation (dotted and dashed lines). The NLO calculations are performed for the fixed, $\alphas=0.2$, and running coupling in the hard coefficients. Two sets of curves are plotted for the LHC $\sqrt{s_{NN}}=\SI{5.02}{TeV}$ and RHIC $\sqrt{s_{NN}}=\SI{200}{GeV}$ energies. Figure from Ref.~\cite{Stasto:2013cha}.}
 \label{fig:scaledependence}
\end{center}
\end{figure}

The reduction of the scale dependence at NLO is illustrated in Fig.~\ref{fig:scaledependence}. As is seen from this figure, the leading order result is quite sensitive to the choice of the  factorization scale $\facscale$. This is understandable as both the parton distribution and fragmentation functions depend on $\facscale$ quite sharply for large values of $\xprojectile > 0.1$ and $\fragfrac > 0.2$. On the other hand, in the NLO calculation the scale dependence is canceled out (up to the one loop order) as is shown in this figure. The calculation also demonstrates that the best choice of the factorization scale is in the region where $\facscale \sim 2-3 \pperp$. 

The most dramatic feature of the NLO calculation, however, is the fact that 
it turns negative at moderate to high transverse momentum, depending on rapidity. This confirms previous calculations~\cite{Altinoluk:2011qy} based on the partial calculation of the NLO contributions.  The NLO correction becomes negative and then it dominates over the LO term at some values of the transverse momentum. The critical value at which the cross section becomes negative depends on the rapidity as can be seen from Fig.~\ref{fig:completenloresults}. The higher the rapidity, the larger the critical value at which the calculation turns negative. 

\begin{figure}
 \centering
 \includegraphics{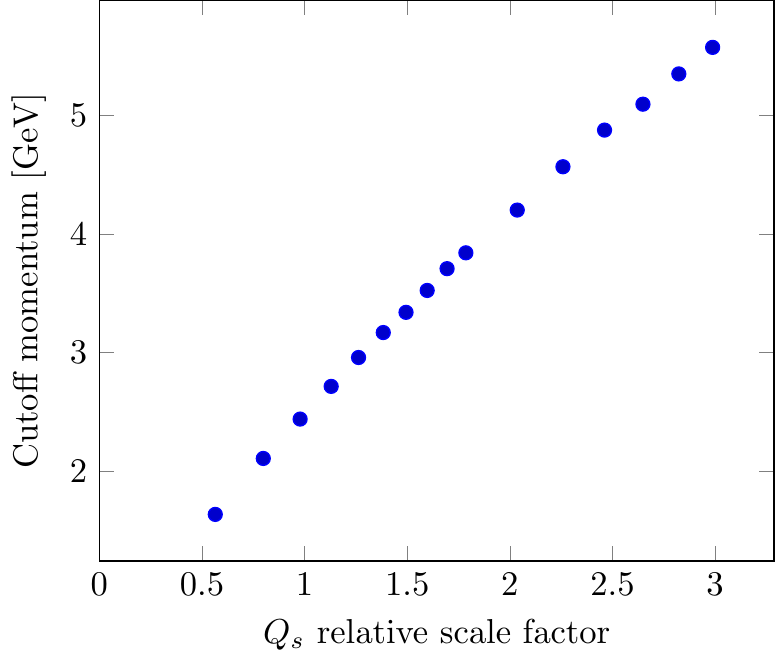}
 \caption{Scaling of the cutoff momentum with the saturation scale, using the LO+NLO cross section with the $L_q$ and $L_g$ corrections included (see Section~\ref{sec:implkc}). Since each calculation of the cross section incorporates a range of values of the saturation scale, we cannot assign a specific value of $\satscale$ to each point. The number on the horizontal axis is an overall fixed scaling factor applied to $\satscale$, such that a scaling factor of $1$ corresponds to minimum bias collisions. }
 \label{fig:cutoff}
\end{figure}

\begin{figure}
\begin{center}
\includegraphics[width=0.85\textwidth]{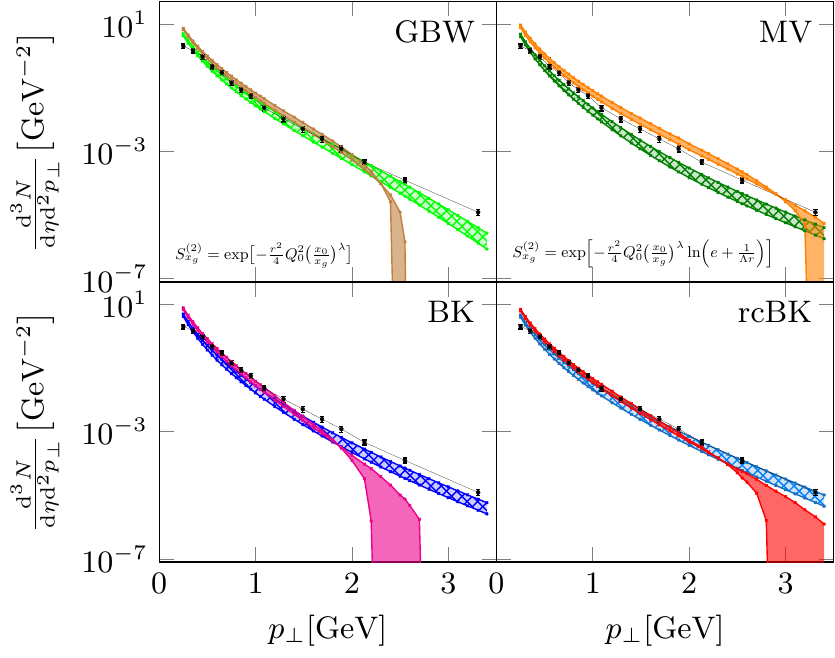}
 \caption{Results of the full NLO calculation (solid bands), from Ref.~\cite{Stasto:2013cha} showing the comparison of the calculations with the experimental data
 from BRAHMS for four different choices of the unintegrated gluon distributions: GBW, MV models, and two solutions to the BK equation with fixed coupling $\alphas=0.1$ and running coupling. The bands correspond to the variation of the scale $\facscale^2=\SI{10}{GeV}^2$ to $\SI{50}{GeV^2}$. The crosshatch fill denotes the LO calculation.}
 \label{fig:gluoncomparison}
 \end{center}
\end{figure}

The existence of this negativity is independent of the form of the gluon distribution. This fact is illustrated in Fig.\ref{fig:gluoncomparison}
where we show the calculations performed with different form of the unintegrated gluon distribution: the GBW model, the MV  model and two versions of the BK equation with the running coupling corrections.
All these calculations agree with the data at low $\pperp$ but then turn negative at high $\pperp$.
The exact cutoff momentum where the negativity sets in depends somewhat on the form of the gluon distribution, but the feature persists in all cases.
As shown in figure~\ref{fig:cutoff}, the cutoff momentum bears an approximately linear relationship to the saturation scale.

In principle, the fact that the cross section is negative does not \emph{necessarily} indicate a problem with the small-$\bjorkenx$ formalism.
This result only reflects the first two terms in a perturbation series, and it's entirely possible that higher-order terms compensate for the negativity, giving a positive, finite result.
However, the result is still somewhat disconcerting.
The fact that the NLO correction is larger than the LO result at high $\pperp$ leads one to wonder whether higher-order terms will continue to grow larger and larger as diagrams with more and more loops are incorporated.
Even though it's not practical to calculate these higher-order contributions (barring some automated method to compute higher-order terms), this would indicate that the perturbation series is divergent and that results up to any finite order cannot be trusted.
In other words, even if we did somehow manage to incorporate the NNLO contribution, it's not clear that it would provide any better predictive value at large $\pperp$ than the current results.
This would leave us largely unable to constrain the gluon distribution at high $\kperp$ using hadron production in $\pA$ collisions.

\subsection{Rapidity Subtraction}\label{sec:rapidity subtraction}

Following the discovery of the negative results, it was quickly established that the negative contribution originates from the plus prescription used in the subtraction of the rapidity divergence, as described in section~\ref{sec:divergences}.
Using the definition of the plus distribution, we can write the finite term from Eq.~\eqref{eq:separate subtraction term} as
\begin{equation}
 \int_a^1 \frac{f(\emitfrac)}{(1 - \emitfrac)_+}\dd\emitfrac = \int_a^1 \frac{f(\emitfrac) - f(1)}{1 - \emitfrac}\dd\emitfrac - \int_0^a\frac{f(1)}{1 - \emitfrac}\dd\emitfrac \; .
\end{equation}
For the relevant functions in the NLO diagonal channels, $f(\emitfrac)$ achieves its maximum value within the range $[0,1]$ at $1$, so $f(\emitfrac) - f(1) \leq 0$.
The negativity in the region near $\emitfrac = 1$ is amplified by the denominator going to zero.

With this in mind, it's natural to consider modifying the subtraction procedure in an attempt to mitigate the negativity of the cross section.
Two recent papers \cite{Kang:2014lha,Ducloue:2016shw} have proposed introducing a cutoff on the momentum fraction, $\rapfac$ (in the notation of Ref.~\cite{Ducloue:2016shw}), which alters the subtraction procedure from Eq.~\eqref{eq:separate subtraction term} by separating the high- and low-momentum gluon emissions as described in section~\ref{sec:rapidity divergences}.
\begin{equation}
  \int^1 \frac{\dd\emitfrac}{1 - \emitfrac}f(\emitfrac) = \int^1 \frac{\dd\emitfrac}{(1 - \emitfrac)_+}f(\emitfrac) + \int_0^{\rapfac} \frac{\dd\emitfrac}{1 - \emitfrac}f(1) + \int_{\rapfac}^1 \frac{\dd\emitfrac}{1 - \emitfrac}f(1)\; .
\end{equation}
With this done, only the last term, representing the low-momentum emissions, is absorbed into the renormalized gluon distribution.
\begin{equation}\label{eq:momentum space modified subtraction}
 \dipoleF[\rapfac](\qperp) \defn \dipoleFbare(\qperp) + \frac{\alphas}{2\pi^2}\int_{\rapfac}^1\frac{\dd\emitfrac}{1 - \emitfrac}\Bigl[S_{qq}^{\text{real}} - S_{qq}^{\text{virt}}\Bigr]_{\emitfrac=1}\; ,
\end{equation}
or, in position space (and using the fact that $\dipoleSbare = \dipoleS[\rapfac]$ to leading order in $\alphas$),
\begin{multline}\label{eq:position space modified subtraction}
 \dipoleS[\rapfac](\vec\xperp, \vec\yperp) = \dipoleSbare(\vec\xperp, \vec\yperp) \\
 - \frac{\alphas\Nc}{2\pi^2}\int_{\rapfac}^1\frac{\dd\emitfrac}{1 - \emitfrac}\int\dd[2]\vec\bperp\frac{(\vec\xperp - \vec\yperp)^2}{(\vec\xperp - \vec\bperp)^2(\vec\yperp - \vec\bperp)^2}\Bigl[\dipoleS[\rapfac](\vec\xperp, \vec\yperp) - \quadrupoleS[\rapfac](\vec\xperp, \vec\bperp, \vec\yperp)\Bigr] \; .
\end{multline}
The cutoff $\rapfac$, or more precisely the logarithm $\ln\frac{1}{1 - \rapfac}$ which is the minimum rapidity of emitted (slow) gluons represented by the subtraction term, provides the evolution variable we need to transform this into the BK equation.

\begin{figure}
 \centering
 \includegraphics[width=0.65\textwidth]{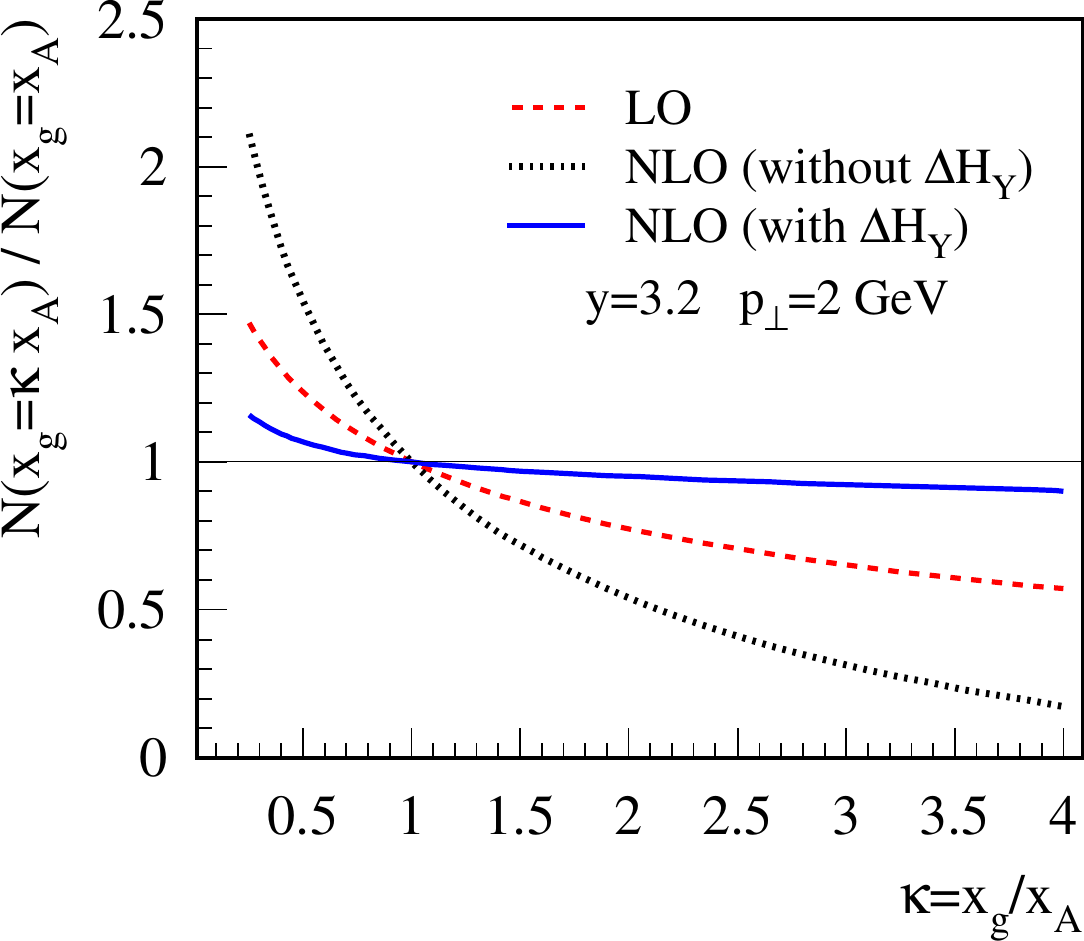}
 \caption{Dependence of the cross section on the rapidity factorization scale using the cutoff chosen in Ref.~\cite{Kang:2014lha}. Reprinted figure with permission from %
 Kang \textit{et al} Phys. Rev. Lett. 113 (2014) 062002; http://dx.doi.org/10.1103/PhysRevLett.113.062002. Copyright 2014 by American Physical Society.}
 \label{fig:rapidity factorization variation}
\end{figure}
\begin{figure}
\begin{center}
 \includegraphics[width=0.48\textwidth]{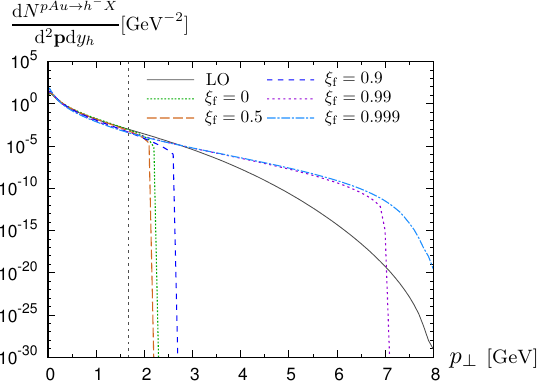}\hspace*{0.5cm}
 \includegraphics[width=0.48\textwidth]{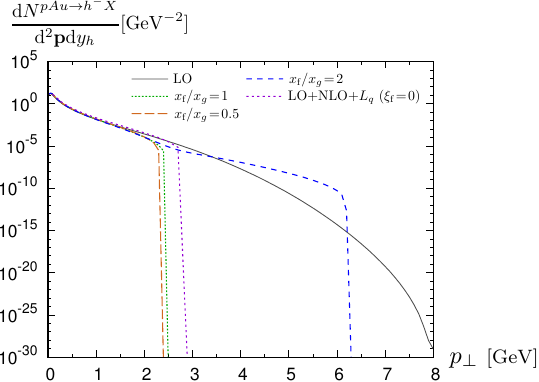}
\end{center}
 \caption{Variation of the cross section with different choices of the rapidity factorization scale $\rapfac$ as computed in Ref.~\cite{Ducloue:2016shw}. The plot on the left shows results using a fixed factorization scale cutoff, while the plot on the right shows the results for momentum-dependent $\rapfac$ as in Eq.~\eqref{eq:k-dependent rapidity cutoff}.
 Reprinted figure with permission from %
 Duclou\'e \textit{et al} Phys. Rev. D93, (2016) 114016; http://dx.doi.org/10.1103/PhysRevD.93.114016. Copyright 2016 by The American Physical Society.}
 \label{fig:rapidity factorization cross section}
\end{figure}
The remaining term, $\int_0^{\rapfac} \frac{f(1)}{1 - \emitfrac}\dd\emitfrac$, is grouped into the finite part of the cross section, unlike Eq.~\eqref{eq:separate subtraction term} where it was considered part of the subtraction term.
This additional finite term becomes a new $\order{\alphas}$ contribution to the cross section, one which is positive at high $\pperp$ and could potentially cancel out the negativity in the original NLO results.
The new contribution takes the form
\begin{align}
 \frac{\dd[3]\sigmainclusive}{\dd\rapidity\dd[2]\pperp}
 &= \frac{\alphas}{2\pi^2}\int_\tau^1 \frac{\dd\fragfrac}{\fragfrac^2}D_{h/q}(\fragfrac) \xprojectile q(\xprojectile) \int_{0}^{\rapfac}\frac{\dd\emitfrac}{1 - \emitfrac}
 \Bigl[S_{qq}^{\text{real}} - S_{qq}^{\text{virt}}\Bigr]_{\emitfrac=1} \\
 &= \frac{\alphas}{2\pi^2}\int_\tau^1 \frac{\dd\fragfrac}{\fragfrac^2}D_{h/q}(\fragfrac) \xprojectile q(\xprojectile) \ln\biggl(\frac{1}{1 - \rapfac}\biggr)
 \Bigl[S_{qq}^{\text{real}} - S_{qq}^{\text{virt}}\Bigr]_{\emitfrac=1}
 \; .
\end{align}
We can see that the value of the cutoff $\rapfac$ affects the calculated NLO cross section.
In fact, the effect can be quite significant, and even brings the cross section from negative to positive over large ranges of $\pperp$.
Given the strong dependence on $\rapfac$, it is important to choose the most sensible value.
Choosing $\rapfac = 0$ reproduces the original result of Ref.~\cite{Chirilli:2012jd}, but more recent work takes varying views on how the value should be fixed.

Kang et al.\cite{Kang:2014lha} argued that $\rapfac$ should be chosen similar to the value of $\xtarget$ to which the gluon distribution $\dipoleS$ is evolved.
Since our renormalization of $\dipoleS$ has incorporated the evolution into the leading order term, we use the leading order kinematics, $\rapfac = \xtarget = \frac{\pperp}{\fragfrac\sqs}e^{-y}$.
However, their results suggest that $\rapfac$ can vary by a factor of order $1$ without changing the cross section too much, as shown in figure~\ref{fig:rapidity factorization variation}.

In a response, Xiao and Yuan~\cite{Xiao:2014uba} have claimed that this rapidity subtraction term is not quite correct because $\rapidity$ should actually be the rapidity difference between the radiated gluon and the projectile \emph{parton} (e.g. quark), which is $\ln\frac{1}{x_g}$, not the projectile hadron (e.g. proton or deuteron).

More recently, Duclou\'e et al.\cite{Ducloue:2016shw} performed a more detailed analysis, showing the effect of various choices for the cutoff~$\rapfac$ over a wider range of possible values.
As shown in the left panel of Fig.~\ref{fig:rapidity factorization cross section}, the effect is very pronounced at high hadron momenta, and the cutoff momentum, at which the LO+NLO cross section becomes negative, varies over nearly the entire kinematically allowed range as $\rapfac$ varies.

Instead of a fixed cutoff value, they propose a momentum-dependent cutoff $\rapfac(\kperp)$, motivated by ordering of the emitted gluons in $k^-$.
Under this scheme, the subtraction term in Eqs.~\eqref{eq:momentum space modified subtraction} and~\eqref{eq:position space modified subtraction} should include gluon emissions in which the fluctuation of the light-cone energy, $\Delta k^-$ (in the notation of our Fig.~\ref{fig:nlo-kinematics}), is at least $x_{\mathrm{f}}p_a^{-}$ for some value $x_{\mathrm{f}}$, which would likely be close to $\xtarget$.
This results in a formula like the following for the cutoff:
\begin{equation}\label{eq:k-dependent rapidity cutoff}
 \rapfac(\kperp) = \frac{\kperp^2}{\kperp^2 + (\xtarget / x_{\mathrm{f}})\satscale^2}\; .
\end{equation}
The associated results are shown in Fig.~\ref{fig:rapidity factorization cross section} on the right.

Each of these methods enhances the cross section at moderate $\pperp$, thus increasing the region in which the LO+NLO result is positive.
However, at sufficiently high momenta, the negativity always comes back.
Section~\ref{sec:expansion} will address the question of whether any prescription can completely cure the negativity at arbitrarily high $\kperp$.

\subsection{Exact kinematics and matching to the collinear calculation}

\begin{figure}
\begin{center}
\includegraphics[width=6.2cm]{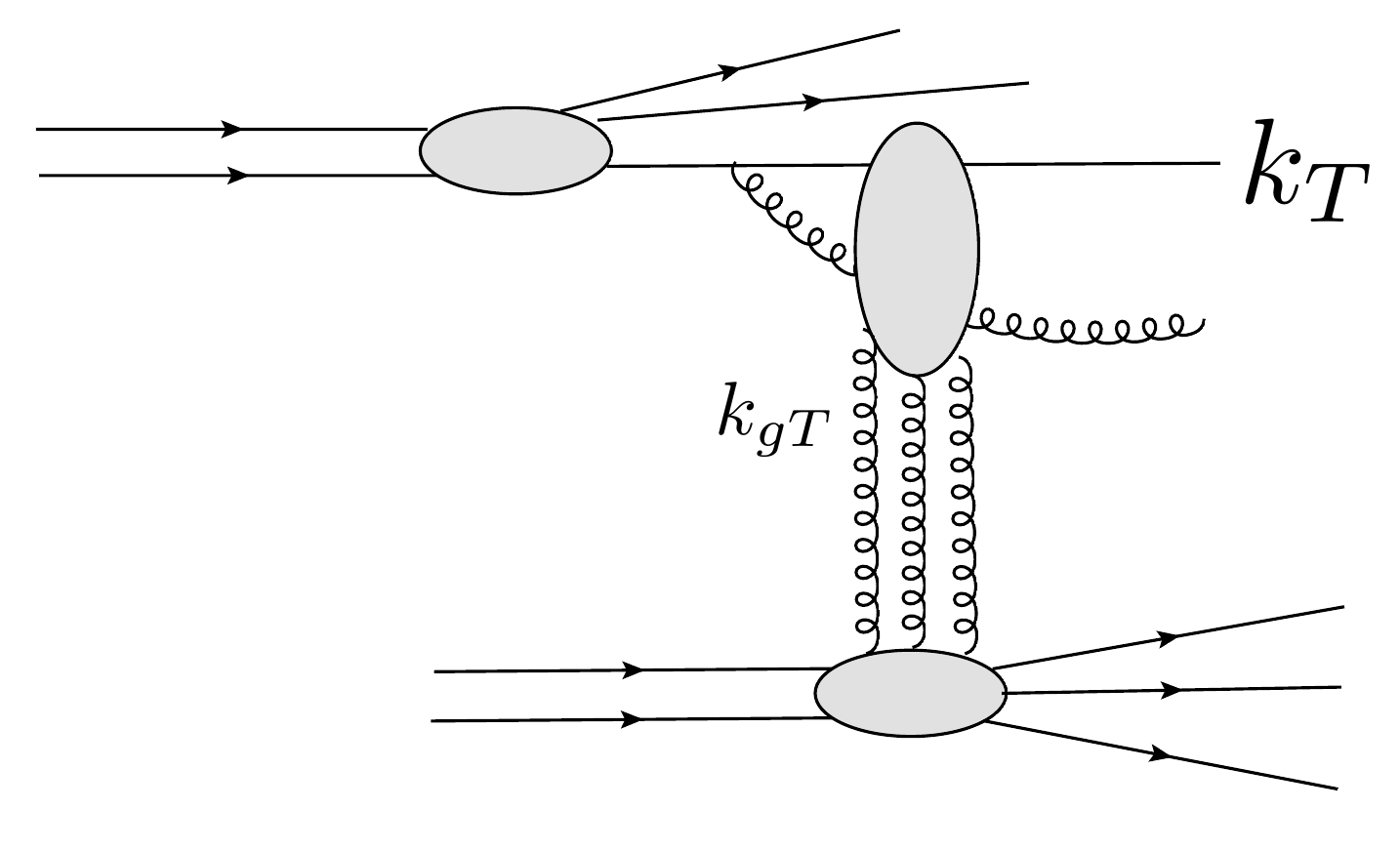}\hspace*{0.8cm}
\includegraphics[width=4.7cm]{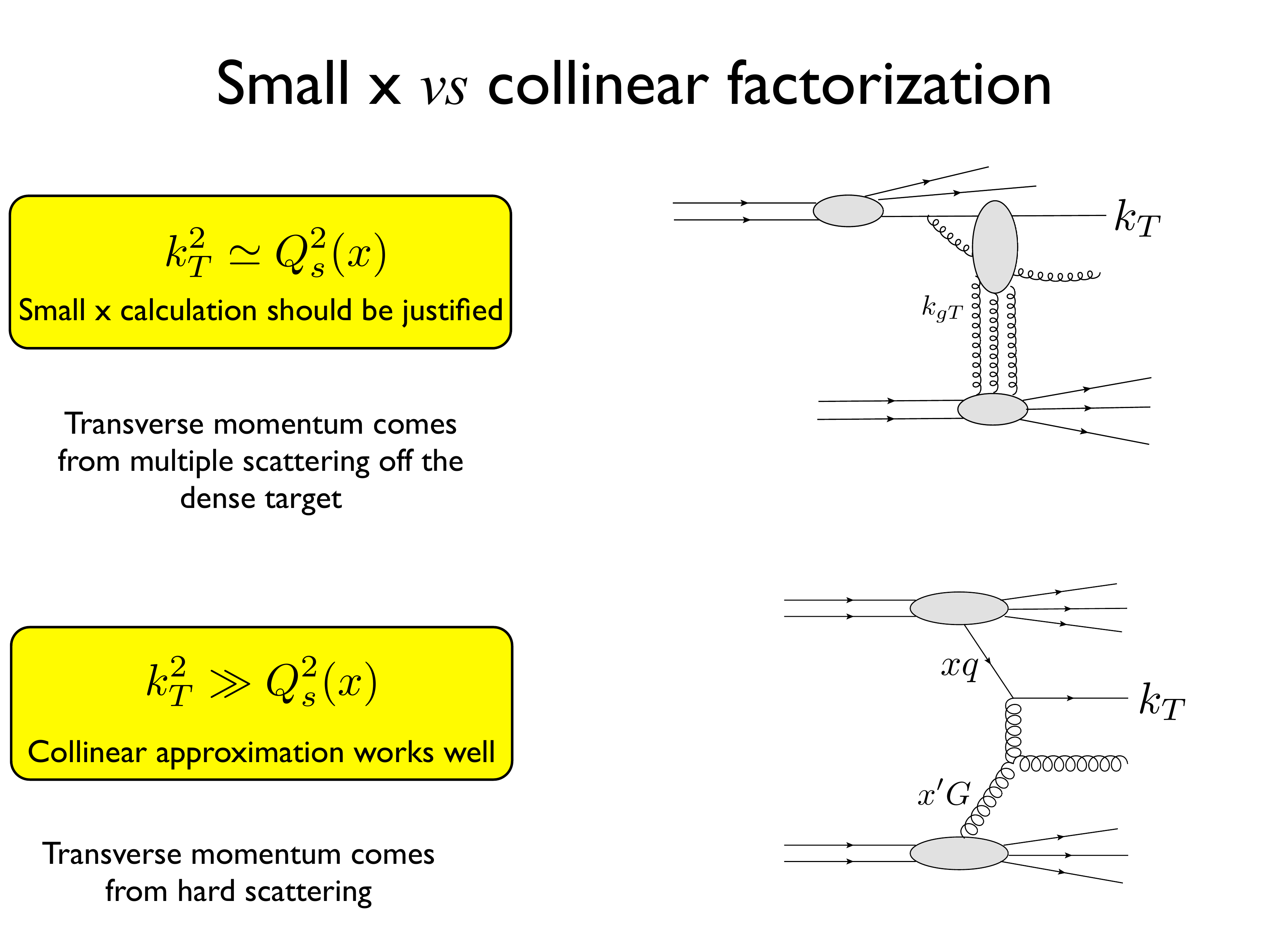}
\end{center}
 \caption{Left: Hybrid calculation at NLO. Right: collinear calculation at LO.}
 \label{fig:collvshybrid}
\end{figure}

The NLO calculation incorporates the $2\to 2$ processes with an off-shell gluon from the target side. In principle, in the collinear approximation, or when the transverse momentum of this gluon is relatively small, this calculation should match into the collinear calculation. We shall emphasize that the origin of the transverse momentum of the final state hadron
is from the hard scattering subprocess only in the collinear factorization and from the transverse momentum of the unintegrated gluon and the hard process in the hybrid approach at NLO. This is schematically illustrated in Fig.~\ref{fig:collvshybrid}.

The matching to the collinear factorization can be shown by expanding the exact NLO formulae (before any subtractions are performed) in powers of $\satscale^2/\kperp^2$ in the large limit of $\kperp^2 \gg \satscale^2$. In principle, systematic expansion leads to the twist expansion, in the spirit of the calculations presented in Refs.~\cite{Bartels:2000hv,Bartels:2009tu}. To match to the collinear calculation, only the leading power in the expansion is retained. In Ref.~\cite{Stasto:2014sea} this expansion was performed both for the $q\to q$ and $g \to g$ channels, with the following formulae for the leading terms:
\begin{multline}
\frac{\dd[3]\sigmainclusive_{q\to q}}{\dd\quarkrapidity \dd[2]\vec\pperp} = \frac{\alphas}{2\pi^{2}}\int_{\tau}^1 \frac{\dd\fragfrac}{\fragfrac^{2}} \fragmentation{q}{h}{\fragfrac}{\facscale^2}\int_{\tau/\fragfrac}^{1}\dd\emitfrac \frac{1+\emitfrac ^{2}}{1-\emitfrac } \partondist{q}{\frac{\xprojectile}{\emitfrac}}{\facscale^2}  \\
\times \biggl\{ \CF\frac{(1-\emitfrac)^2}{\kperp^4} + \Nc\frac{\emitfrac}{\kperp^4}\biggr\}
\int_{\mathcal{R}} \dd[2] \vec\qperp \qperp^2\dipoleF(\qperp)\, , 
\end{multline}
\begin{multline}
\frac{\dd[3]\sigmainclusive_{g\to g}}{\dd\gluonrapidity \dd[2]\vec\pperp}  =  \frac{ \Nc }{2\pi^2}\int^1_{\tau} \frac{\dd\fragfrac}{\fragfrac^2} \fragmentation{g}{h}{\fragfrac}{\facscale^2}\int_{\tau/\fragfrac}^{1}\dd\emitfrac \partondist{g}{\frac{\xprojectile}{\emitfrac}}{\facscale^2} \\
\times \frac{2 [1-\emitfrac (1-\emitfrac)]^2 [1+\emitfrac^2+(1-\emitfrac)^2]}{ \emitfrac (1-\emitfrac)}
\frac{1}{\kperp^4}
\int_{\mathcal{R}} \dd[2] \vec\qperp \qperp^2\dipoleF(\qperp).\label{eq:collinearexpansion}
\end{multline}

The integrals over the unintegrated gluon distributions can be then simplified to the integrated, collinear densities in the following way
\begin{equation}
\int_{\mathcal{R}}  \dd[2] \vec\qperp \qperp^2\dipoleF(\qperp) \simeq \frac{2\pi^2}{\Nc} \partondist{g_A}{x'}{\facscale^2} \; ,
\label{eq:unintegrated}
\end{equation}
where $\facscale$ is the scale of the integrated distribution of the nucleus. In this formalism it is set to be equal to $\satscale$.

In order to match to the collinear calculation one needs to carefully  evaluate the kinematics. The exact kinematics for the $2\to 2$ process with energy-momentum conservation is defined by 
\begin{align}
 \xprojectile &= \frac{\kperp}{\sqs\emitfrac}e^{\hadronrapidity}\, , \notag \\
 \xtarget &= \frac{\kperp}{\sqs}e^{-\hadronrapidity} + \frac{(\vec{k}_{g\perp}-\vec{k}_{\perp})^2}{\sqrt{s} k_{\perp}} \frac{\emitfrac}{1-\emitfrac} e^{-y}\label{eq:xa}\, .
\end{align}
The small-$\bjorkenx$ limit requires the center-of-mass energy to be very large, $\mandelstams\rightarrow \infty$, and at the same time the $\xprojectile$ is kept large which corresponds to the forward limit of the hadron production. However, in any practical calculations, even in the LHC kinematics the energy is not that large, and one has to keep kinematics exact. In the small $x$ limit, one takes the gluon transverse momentum to be of the order of the $k_T$, which results in the approximation to the $x_a\simeq x_{g0}$, because
the second term in the above equation is small. However, in the collinear limit this is no longer the case and we have $k_T \gg k_{gT}$, which leads to the following approximation for the gluon longitudinal momentum fraction
\begin{equation}
x' \;  = \;   \frac{k_{\perp}}{\sqrt{s}}e^{-y} + \frac{{k}_{\perp}}{\sqrt{s}} \frac{\emitfrac}{1-\emitfrac} e^{-y}\, .\label{eq:xprim}
\end{equation}

As a result, \cite{Stasto:2014sea} using ``exact kinematics'' (i.e. $x_a<1$ with definition \eqref{eq:xa}) reveals that the largest kinematically allowed value of $\emitfrac$ is
\begin{equation}\label{eq:ximax}
 \emitfrac_\text{max} = \frac{1 - \xglo}{1 - \xglo + \xglo(\vec\kgperp - \vec\kperp)^2/\kperp^2} \; ,
\end{equation}
where $\xglo = \frac{\kperp}{\sqs}e^{-\rapidity}$ is the definition of the gluon momentum fraction in leading order kinematics (although in this formula it is simply a convenient abbreviation and does not represent a physical momentum fraction).
This is strictly less then $1$ except when $\vec\kgperp = \vec\kperp$, i.e. when the emitted gluon has zero transverse momentum.

Going back to formula \eqref{eq:collinearexpansion} and using \eqref{eq:unintegrated} together with the collinear kinematics \eqref{eq:xprim} one can demonstrate that indeed it coincides with the leading order collinear formula.

\begin{figure}
 \centering
 \includegraphics[width=0.85\textwidth]{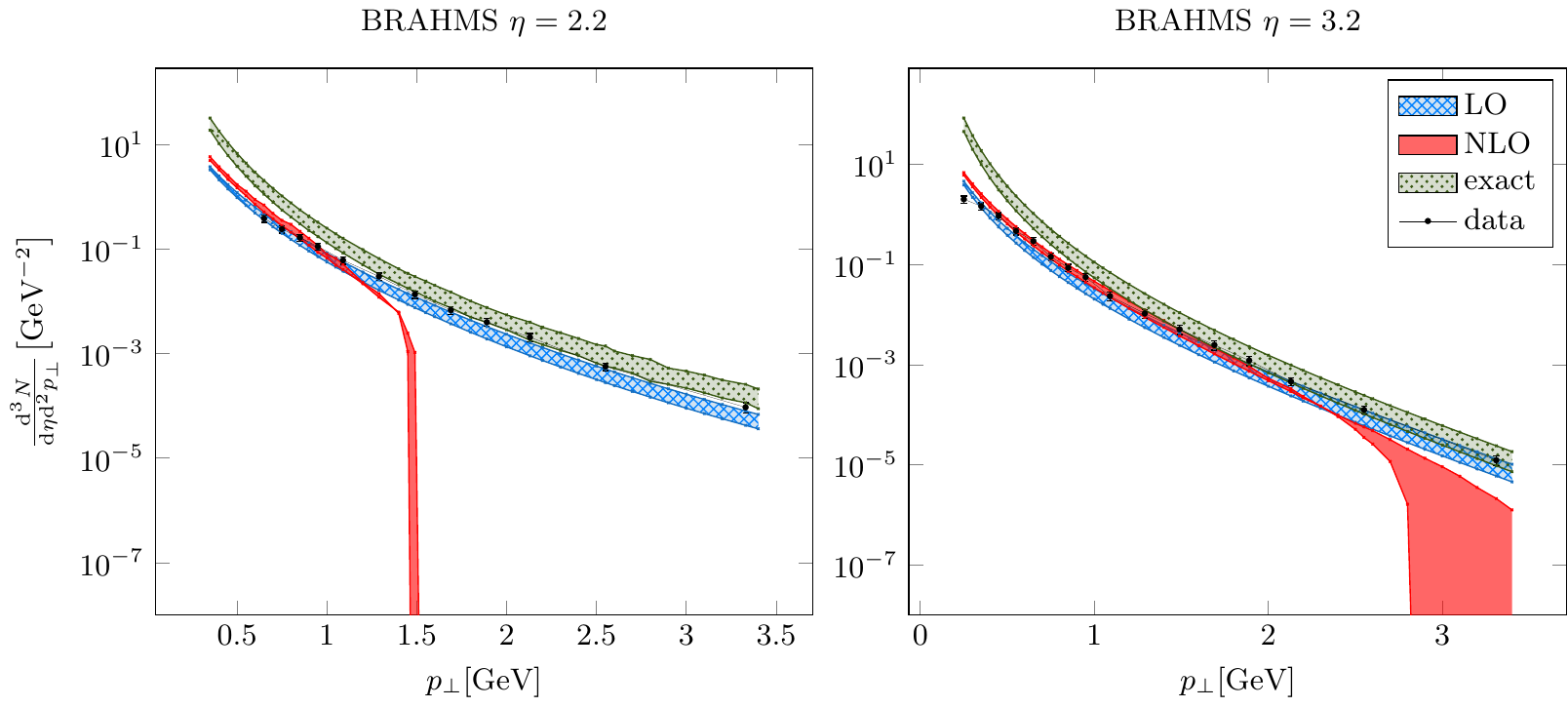}
 \caption{Comparison of the LO, NLO small $x$-calculations~\cite{Stasto:2014sea} together with the leading power expansion with exact kinematics at rapidities $\hadronrapidity=2.2,3.2$ at $\sqsnn=\SI{200}{GeV}$. The data are from BRAHMS~\cite{Arsene:2004ux}, the calculations use running coupling BK equation. The figure is taken  from Ref.~\cite{Stasto:2014sea}.}
 \label{fig:collinearmatching}
\end{figure}

One can extract the dominant contributions to the cross section at high $\kperp$ in momentum space and evaluate those with the constraint in force.
The result~\cite{Stasto:2014sea} is positive and matches experimental data fairly well at high $\pperp$.
The comparison of the LO and NLO calculations is shown in Fig.~\ref{fig:collinearmatching}. One can see that the calculation with the expansion of the kinematics coincides well with the data at large $\pperp$ and stays positive. On the other hand it does also match to the NLO calculation for intermediate values of the $\pperp$, of the order of the saturation scale. For lower values of $\pperp$, the expansion is overshooting the NLO calculation, which includes more of the higher twist effects in this region and better matches the experimental data.

Furthermore, at high transverse momentum, $\kperp\gg\satscale$ so saturation effects are generally negligible. 
Therefore the perturbative description (which does not account for the nonlinear phenomenon of saturation) should be accurate, and indeed the high-$\kperp$ approximation can be analytically shown to coincide with the result from collinear factorization and perturbative QCD at large $\pperp$.

\subsection{Implementation of the kinematical constraint}
\label{sec:implkc}
\begin{figure}
\begin{center}
 \includegraphics[width=0.85\textwidth]{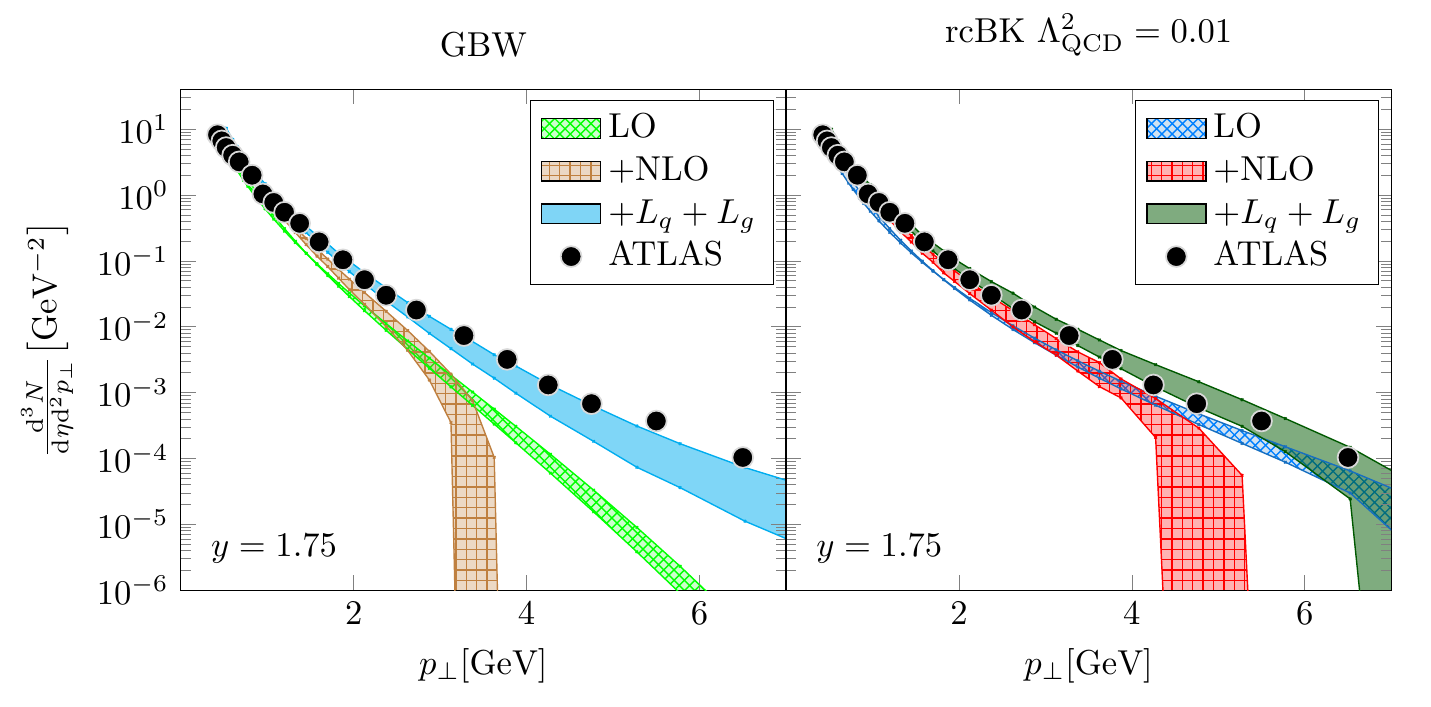}
 \caption{Contribution of $L_q$ and $L_g$ to the cross section, equations~\eqref{eq:Lq cross section} and~\eqref{eq:Lg cross section}, along with the leading and original next-to-leading contributions for reference. The contribution from $L_q$ and $L_g$ is able to restore the cross section to a positive result up to moderate $\pperp$. Data are from ATLAS with center of mass energy $\sqsnn=\SI{5.02}{TeV}$ compared with the SOLO results for the GBW model and the running coupling BK. Figure referenced from \cite{Watanabe:2015tja}.}
 \label{fig:lqlgresults}
 \end{center}
\end{figure}

There has been recent progress\cite{Altinoluk:2014eka,Watanabe:2015tja} towards evaluating the cross section subject to the constraint~\eqref{eq:ximax}.
Two groups take different approaches.

The work of Altinoluk et al.\cite{Altinoluk:2014eka} describes the Ioffe time restriction of the split pair.
Their argument is as follows: when a projectile parton emits a gluon prior to interacting with the target nucleus, the resulting pair has a coherence time
\begin{equation}
 \coherencetime \sim \frac{2\emitfrac(1 - \emitfrac)\xprojectile P^+}{\kperp^2} \; ,
\end{equation}
where it is assumed that the final-state gluon and progenitor parton have equal and opposite transverse momenta $\pm\vec\kperp$.
If $\coherencetime$ for a given $qg$ pair is less than the time $\targettime$ it takes to traverse the target, the pair behaves as a single dressed quark, not as a resolved parton pair.
Therefore, when accounting for gluon emission at NLO, we should omit the region of phase space in which $\coherencetime < \tau$.
This leads to the kinematic constraint
\begin{equation}
 \frac{\emitfrac(1 - \emitfrac)\xprojectile}{\kperp^2} > \frac{1}{\mandelstams} \; .
\end{equation}
Since this constraint is relevant only for $\emitfrac \approx 1$, we can approximate it as
\begin{equation}
 \emitfrac \lesssim 1 - \frac{\kperp^2}{\xprojectile \mandelstams} \; .
\end{equation}
Application of the constraint then leads\cite{Altinoluk:2014eka} to a modification of the Weisz\"acker-Williams field; specifically, it introduces a factor of
\begin{equation}\label{eq:ioffe time constraint}
 1 - \besselJzero\Biggl(\uperp\sqrt{2\emitfrac(1 - \emitfrac)\frac{\xprojectile P^+}{\tau}}\Biggr)\,,
\end{equation}
where $\vec\uperp = \vec\xperp - \vec\bperp$ is the transverse separation of the original dipole, due to the restriction imposed by the constraint on the Fourier transform.

Alternatively, Watanabe et al.\cite{Watanabe:2015tja} justify the constraint using conservation of the minus component of four-momentum.
\begin{equation}
\xtarget = \frac{\qperp^2}{(1 - \emitfrac)\xprojectile \mandelstams} + \frac{\kperp^2}{\emitfrac \xprojectile \mandelstams} \leq 1\, .
\end{equation}
Here $\qperp$ is the gluon momentum and $\kperp$ is that of the progenitor parton.
As $\emitfrac\to 1$, the first term becomes dominant and this becomes
\begin{equation}
 \emitfrac \lesssim 1 - \frac{\qperp^2}{\xprojectile \mandelstams} \; .
\end{equation}
This constraint can be applied to the dipole splitting function, which introduces a factor of
\begin{equation}
 1 - \besselJzero\Biggl(\uperp\sqrt{\emitfrac(1 - \emitfrac)\xprojectile \mandelstams}\Biggl) \; ,
\end{equation}
again arising from the constraint acting on the Fourier transform.
This is equivalent to equation~\eqref{eq:ioffe time constraint} if one takes $\frac{2P^+}{\targettime} = \mandelstams$.

The final result of the constraint is an additional contribution to the cross section which can be broken down into two terms, one from the quark-quark channel and another from the gluon-gluon channel.
Respectively,
\begin{align}
 \frac{\dd[3]\sigma_{L_q}}{\dd\rapidity\dd[2]\pperp} &= \int_\tau^1 \frac{\dd\fragfrac}{\fragfrac^2}\sum_f \partondist{q_f}{\xprojectile}{\facscale^2}\fragmentation{f}{h}{\fragfrac}{\facscale^2} L_q(\kperp) \; ,\label{eq:Lq cross section} \\
 \frac{\dd[3]\sigma_{L_g}}{\dd\rapidity\dd[2]\pperp} &= \int_\tau^1 \frac{\dd\fragfrac}{\fragfrac^2} \partondist{g}{\xprojectile}{\facscale^2}\fragmentation{g}{h}{\fragfrac}{\facscale^2} L_g(\kperp) \; . \label{eq:Lg cross section} 
\end{align}
Explicit expressions for the functions $L_q$ and $L_g$ can be found in Watanabe et al.\cite{Watanabe:2015tja}

Numerically, one finds that these kinematical correction terms are positive and large enough to restore the positivity of the cross section at intermediate $\pperp$.
Figure~\ref{fig:lqlgresults} shows that the cutoff momentum at which the cross section becomes negative is larger with the correction.

\subsection{Expansion and identification of negativity}\label{sec:expansion}

\begin{figure}
 \centering
 \includegraphics{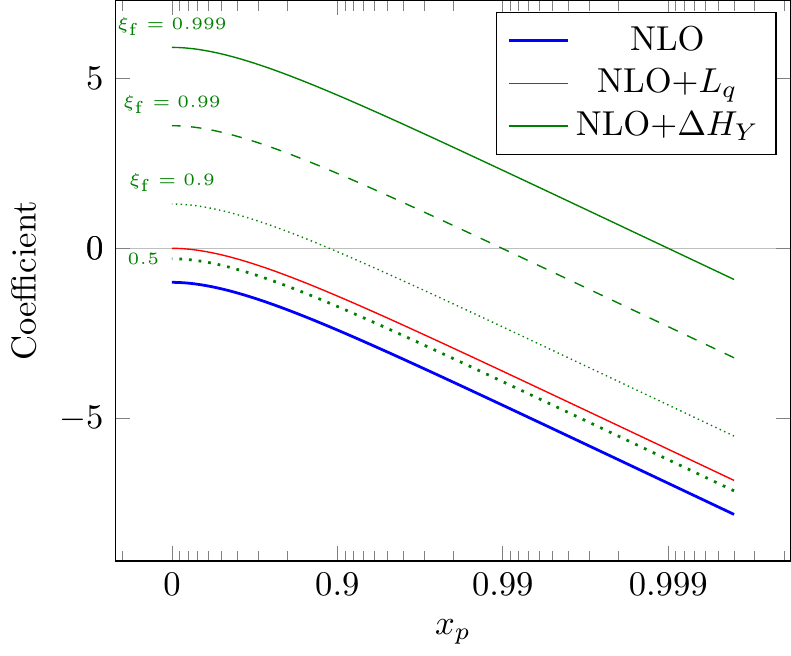}
 \caption{The lowest curve, labeled ``NLO'', shows the negative coefficient in the high-$\kperp$ expansion of the cross section, $\ln(1 - \xprojectile) - (1 - \xprojectile)$ from Eq.~\eqref{eq:lo+nlo asymptotic xp expansion}. The higher curves show how this changes with the addition of the $L_q$ term~\eqref{eq:Lq asymptotic} or the rapidity subtraction correction~\eqref{eq:HY asymptotic}.}
 \label{fig:asymptotic-coefficient}
\end{figure}

As previously discussed, there have been several proposals \cite{Kang:2014lha,Ducloue:2016shw,Watanabe:2015tja,Altinoluk:2014eka} for modifications to the NLO cross section which address the negativity, but so far none appear to cure it entirely.
They simply push the cutoff momentum at which the cross section turns negative up to higher values.
A better solution would truly remove the negativity from the result at all values of $\pperp$.
In theory, this should be doable with an all-order resummation, but developing a suitable resummation procedure is quite difficult and there has not been useful progress in this area.
If the negativity can be truly cured with fixed-order terms, it could represent a significant improvement in the predictive value of the formula.

Since the negativity is strongest at high momenta, let's examine the high-momentum limit of the original LO+NLO cross section derived by Chirilli et al.\cite{Chirilli:2012jd}.
We will limit our calculation to the quark-quark channel, which is expected to be representative of the full cross section.
Expanding the integrand in powers of $\frac{\satscale^2}{\kperp^2}$, we find that the leading contribution to the integrand is of order $\order{\kperp^{-4}}$ and takes the form
\begin{equation}\label{eq:lo+nlo asymptotic}
 \frac{\alphas}{2\pi^2} \int_\tau^1\dd\fragfrac \int_{\tau/\fragfrac}^1\dd\emitfrac\ \frac{\partondist{q}{\frac{\xprojectile}{\emitfrac}}{\facscale^2}\fragmentation{h}{q}{\fragfrac}{\facscale^2}}{\fragfrac^2}\frac{1 + \emitfrac^2}{(1 - \emitfrac)_+}\Bigl(\CF(1 - \emitfrac)^2 + \Nc\emitfrac\Bigr) \frac{2\pi}{\kperp^4}\int\dd[2]\vec\qperp\,\qperp^2\dipoleF(\qperp)
\end{equation}
To compare this to the other contributions, we need to integrate over $\emitfrac$.
Although we can't do the integral exactly without an explicit form for $\partondist{q}{\frac{\xprojectile}{\emitfrac}}{\facscale^2}$, we can take the leading term in a series expansion around $\frac{\tau}{\fragfrac} = \xprojectile \approx 1$.
We expect this to be a reasonable approximation because $\tau \leq \emitfrac \leq 1$, and $\tau \propto \pperp$, so when $\pperp$ becomes large, the range of allowed values for $\emitfrac$ is small.
When doing the $\emitfrac$ integral, the term proportional to $\CF$ vanishes, and we obtain
\begin{multline}\label{eq:lo+nlo asymptotic xp expansion}
 \frac{\alphas\Nc}{2\pi^2} \int_\tau^1\dd\fragfrac \frac{\partondist{q}{\xprojectile}{\facscale^2}\fragmentation{h}{q}{\fragfrac}{\facscale^2}}{\fragfrac^2} \bigl(\ln(1 - \xprojectile) - (1 - \xprojectile)\bigr) \frac{4\pi}{\kperp^4}\int\dd[2]\vec\qperp\,\qperp^2\dipoleF(\qperp)
 \\ +
 \frac{\alphas\Nc}{2\pi^2} \int_\tau^1\dd\fragfrac \frac{\xprojectile^2 q'\left(\xprojectile, \facscale^2\right)\fragmentation{h}{q}{\fragfrac}{\facscale^2}}{\fragfrac^2} (1 - \xprojectile) \frac{4\pi}{\kperp^4}\int\dd[2]\vec\qperp\,\qperp^2\dipoleF(\qperp)
\end{multline}
Within the first line, the negativity comes from the factor $\ln(1 - \xprojectile) - (1 - \xprojectile)$, in particular the logarithm, which dominates over the other terms as $\xprojectile\to 1$.\footnote{Note that the integral is not divergent. One can show this by expressing the rest of the integrand as a power series in $\fragfrac$, then using
\begin{equation*}
 \int_\tau^1 \fragfrac^n [\ln(1 - \tau/\fragfrac) - (1 - \tau/\fragfrac)]\dd \fragfrac = (1 - \tau)[\ln(1 - \tau) - 1] + \order{(1 - \tau)^2}\;,
\end{equation*}
which is finite and negative and goes to zero as $\tau\to 1$.}
Any correction term which is to cure the negativity will have to cancel this logarithm.
Meanwhile, on the second line, we see the derivative of the quark distribution $q'(\xprojectile, \facscale^2) = \pdv{\xprojectile}q(\xprojectile, \facscale^2)$.
In the kinematic region we're looking at, the parton distribution decreases as $\xprojectile \to 1$ from below, so we can expect that this contribution will also be negative, although not divergent.

Moving on to the $L_q$ term arising from the kinematic constraint\cite{Altinoluk:2014eka,Watanabe:2015tja}, we can expand it in $\kperp$ and find that it makes the following contribution to the cross section in the high-$\kperp$ limit:
\begin{equation}\label{eq:Lq asymptotic}
 \frac{\alphas\Nc}{2\pi^2} \int_\tau^1 \dd\fragfrac \frac{\partondist{q}{\xprojectile}{\facscale^2}\fragmentation{q}{h}{\fragfrac}{\facscale^2}}{\fragfrac^2} \frac{4\pi}{\kperp^4} \int\dd[2]\vec\qperp\,\qperp^2\dipoleF(\qperp)
\end{equation}
This affects the negative factor from the LO+NLO term~\eqref{eq:lo+nlo asymptotic xp expansion} only by adding a constant.
\begin{equation}
 \underbrace{\ln(1 - \xprojectile) - (1 - \xprojectile)}_{\text{LO+NLO}} + \underbrace{1}_{L_q} = \ln(1 - \xprojectile) + \xprojectile
\end{equation}
As $\xprojectile\to 1$, this is still dominated by the negative logarithm, as shown in Fig.~\ref{fig:asymptotic-coefficient}.

\begin{figure}
 \centering
 \includegraphics[width=0.75\textwidth]{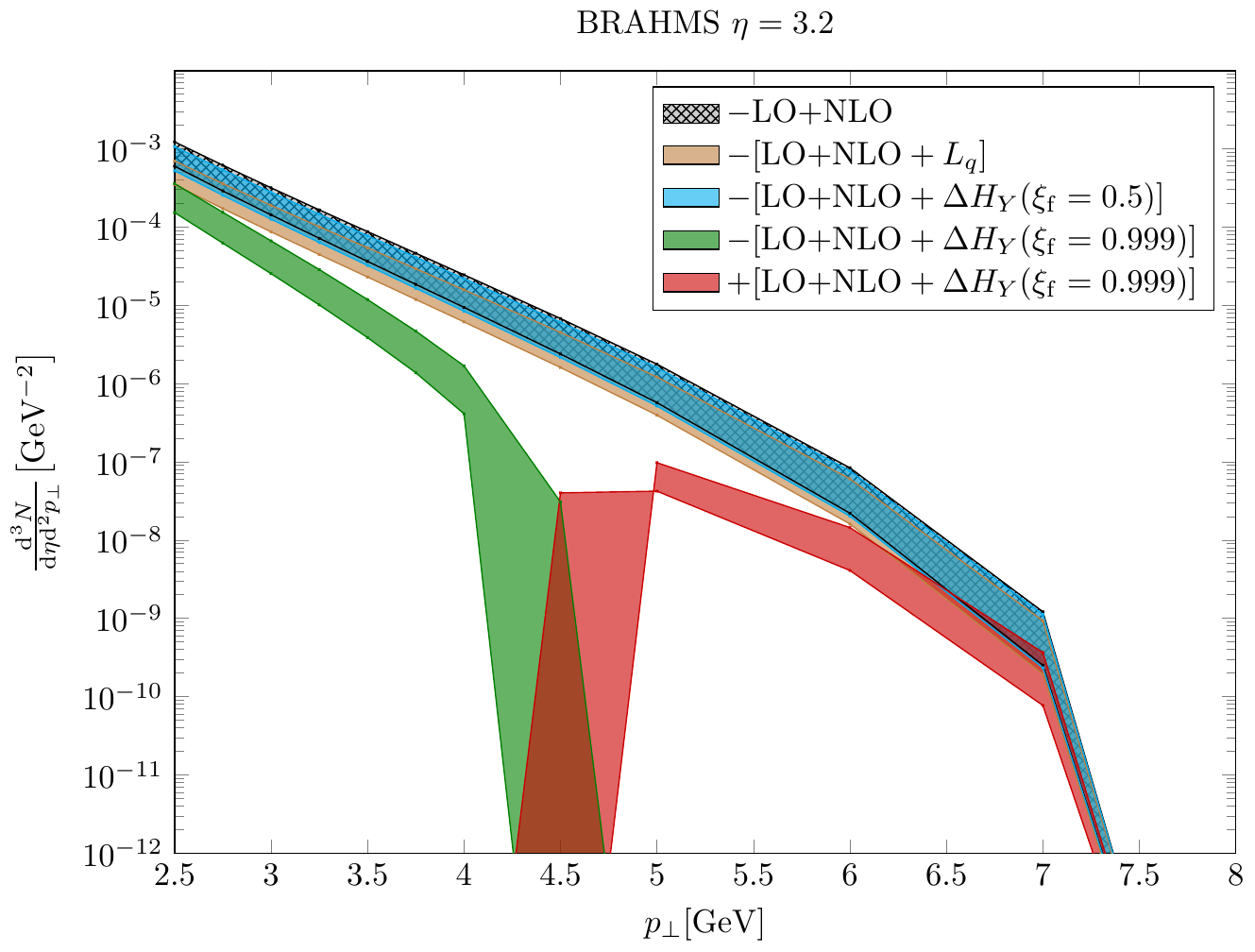}
 \caption{This figure shows the high-$\kperp$ approximation to the differential yield resulting from the original LO+NLO calculation~\cite{Chirilli:2012jd,Stasto:2014sea}, the same quantity with the $L_q$ addition~\cite{Altinoluk:2014eka,Watanabe:2015tja}, and the LO+NLO result with the rapidity subtraction correction~\cite{Ducloue:2016shw,Kang:2014lha} for two different fixed values of $\rapfac$. Except for the latter, all the results are actually negative, so what is plotted here is the absolute value of the yield. The exception is the $\rapfac = 0.999$ curve, which is positive up to $\pperp\approx \SI{4.5}{GeV}$ and negative above, so this result is plotted using two colors, green for the positive part and red for the negative part. The results show clearly the dominance of the negative logarithm at the highest values of $\pperp$. The kinematic limit for these conditions, namely BRAHMS $\sqs = \SI{200}{GeV}$ and $\hadronrapidity = 3.2$, is $\pperp < \SI{8.15}{GeV}$.}
 \label{fig:nlo-xsec-correction}
\end{figure}

If we expand the rapidity subtraction correction~\cite{Ducloue:2016shw,Kang:2014lha}, denoted $\Delta H_Y$, in the same way, we get the contribution
\begin{equation}\label{eq:HY asymptotic}
 \frac{\alphas \Nc}{2\pi^2}\int_{\tau}^1 \dd\fragfrac \frac{\partondist{q}{\xprojectile}{\facscale^2}\fragmentation{q}{h}{\fragfrac}{\facscale^2}}{\fragfrac^2}
 \ln\biggl(\frac{1}{1 - \rapfac}\biggr) \frac{4\pi}{\kperp^4} \int\dd[2]\vec\qperp\,\qperp^2\dipoleF(\qperp)
\end{equation}
This in turn affects the negative factor from Eq.~\eqref{eq:lo+nlo asymptotic xp expansion} as
\begin{equation}
 \underbrace{\ln(1 - \xprojectile) - (1 - \xprojectile)}_{\text{LO+NLO}} + \underbrace{\ln\biggl(\frac{1}{1 - \rapfac}\biggr)}_{\Delta H_Y} = \ln(1 - \xprojectile) + \text{constant}
\end{equation}
The additional term is, again, just a constant with respect to $\xprojectile$.
Although the constant can be made as large as desired by adjusting $\rapfac$, there will always be a small range of $\xprojectile$ close to $1$ where the $\ln(1 - \xprojectile)$ term still dominates, as shown in Fig.~\ref{fig:asymptotic-coefficient}.
Thus there will always be some value of $\pperp$ which puts $\tau$ close enough to $1$ to make the cross section negative even with the $\Delta H_Y$ correction.
Fig.~\ref{fig:nlo-xsec-correction} shows a sample calculation, illustrating how even $\rapfac=0.999$ is not sufficient to cancel the negativity at very high $\pperp$, close to the kinematic limit of $\SI{8.15}{GeV}$.
Of course, one could go to larger and larger values of $\rapfac$, and eventually bring the cross section positive up to some $\pperp$ that is practically indistinguishable from the kinematic limit, so the rapidity subtraction correction is at least a useful phenomenological tool.

Refs.~\cite{Ducloue:2016shw,Kang:2014lha} additionally propose methods of setting the cutoff $\rapfac$ to a momentum-dependent value, which is more general than the constant cutoff considered here.
However, this does not seem likely to change the qualitative result that the negativity persists at very high $\pperp$.
We leave a detailed verification of this fact, as well as investigation of other modification schemes which might be able to cancel out the negative logarithm, to future work.

\section{Summary}

In this review we have briefly summarized recent progress in the calculation of the single inclusive hadron production at forward rapidities within the saturation formalism, as well as its application to phenomenology. This process is particularly useful for testing the small-$\bjorkenx$ dynamics, due to the low values of the longitudinal momentum fraction being probed in the target hadron or nucleus. Thus, it has been applied not only to proton-proton collisions, but also to proton-nucleus collisions in the proton's forward rapidity range.  This formalism, sometimes called the hybrid formalism, employs a combination of the collinear parton distribution functions from the projectile side and the unintegrated parton distribution functions on the target side. The formalism in the lowest order has been very successful in the description of experimental data.

The addition of NLO corrections was a significant step forward to extend the saturation formalism for this process, and it was demonstrated that the factorization still holds at this level. The appropriate divergences, i.e. collinear and rapidity divergences, have been incorporated into the  integrated parton distribution functions of the projectile, the fragmentation functions of the produced hadron, and the unintegrated gluon distribution of the target. Numerical evaluation for this process showed that at this order, the differential distribution fits better to the experimental data at very low transverse momenta and has smaller scale dependence. However, the calculation turns negative at larger values of transverse momenta. Although the precise nature of the negativity, including the transverse momentum at which it sets in, depends on the kinematics, e.g. being more prevalent at lower rapidity, its existence is ``universal'', being independent of the form of the unintegrated distribution used and other parameters. The negativity can be traced to the subtraction of the rapidity divergence through the plus prescription.

Recent work has focused on several paths to remedy this problem. 
Improvements of the kinematics, essentially based on the Ioffe time constraint, have been considered. This improvement generates additional terms in the NLO formalism, which shrink the kinematic range in which the results are negative. Another approach proposes a modification of the rapidity subtraction, and by varying this cutoff one can push the negativity to yet higher values of the transverse momenta. However, we have shown that none of these approaches appear to be capable of eliminating the negativity entirely.

In the future, more improvements to this formalism could be done, among them the calculation using the solution to the nonlinear evolution in the NLL level, or better yet, the resummed version of the nonlinear evolution equation. It would also be interesting to see whether the hybrid form of the factorization can be extended beyond the NLO level.

\section*{Acknowledgments}
This work was supported by the Department of Energy  Grant  No. DE-SC-0002145 and  by
the National Science Center, Poland, Grant No. 2015/17/B/ST2/01838. We would like to thank Francois Arleo, Bertrand Duclou\'e, Zefang (Jimmy) Jiang, Tuomas Lappi, Lech Szymanowski, Bowen Xiao, and Yan Zhu for useful discussions.

\bibliography{nlosatrev}

\end{document}